%% file: TOI-1260.tex
\documentclass[fleqn,usenatbib]{mnras}

\usepackage{newtxtext,newtxmath}

\usepackage[T1]{fontenc}

\usepackage{caption}
\usepackage{subcaption}
\usepackage{threeparttable}
\usepackage{rotating}
\usepackage{rotfloat}
\usepackage{float}
\usepackage{moresize}
\usepackage{wrapfig}
\usepackage{multicol}
\usepackage{wrapfig}
\usepackage[normalem]{ulem}

\usepackage{graphicx}	
\usepackage{amsmath}	
\usepackage{epstopdf}   

\newcommand{\target}{TOI-1260}
\newcommand{\planetb}{TOI-1260\,b}
\newcommand{\planetc}{TOI-1260\,c}
\newcommand{\smw}{S-index}
\newcommand{\fatm}{$f_{\rm atm}$}
\newcommand{\renv}{$R_{\rm env}$}
\newcommand{\msun}{$M_{\odot}$}
\newcommand{\rsun}{$R_{\odot}$}
\newcommand{\lsun}{$L_{\odot}$}
\newcommand{\mearth}{$M_{\oplus}$}
\newcommand{\rearth}{$R_{\oplus}$}
\newcommand{\mstar}{\ensuremath{M_{\star}}}
\newcommand{\rstar}{\ensuremath{R_{\star}}}

\newcommand{\feh}{\ensuremath{[\mbox{Fe}/\mbox{H}]}}
\newcommand{\teff}{\ensuremath{T_{\mathrm{eff}}}}
\newcommand{\logg}{\ensuremath{\log g}}
\newcommand{\vsini}{\ensuremath{V \sin i_\star}}

\newcommand{\mps}{$\mathrm{m\,s^{-1}}$}
\newcommand{\kmps}{$\mathrm{km\,s^{-1}}$}
\newcommand{\rhostar}{\ensuremath{\rho_{\star}}}
\newcommand{\rp}{\ensuremath{R_\mathrm{p}}}
\newcommand{\Mp}{\ensuremath{M_\mathrm{p}}}

\newcommand{\tess}{\textit{TESS}}

\newcommand{\gaia}{\textit{Gaia}}

\newcommand{\stype}{K6\,V} 
\newcommand{\massstype}{0.66} 
\newcommand{\radiusstype}{0.65} 

\newcommand{\steffsme}[1][]{$4227 \pm 85$} 
\newcommand{\sloggsme}{$4.57 \pm 0.05$} 
\newcommand{\sfehsme}[1][]{$-0.10 \pm 0.07$} 
\newcommand{\svsinisme}[1][]{$1.5 \pm 0.7$} 
\newcommand{\svmicsme}[1][]{$1.0$} 
\newcommand{\svmacsme}[1][]{$1.5$} 
 
\newcommand{\steffspecm}[1][]{$4207 \pm 70$} 
\newcommand{\sfehpecm}[1][]{$-0.06 \pm 0.12$} 
\newcommand{\sradiusspecm}[1][]{$0.67 \pm 0.07$} 

\newcommand{\smassisochrones}{$0.66\pm 0.01$}  
\newcommand{\sradiusisochrones}{$0.65\pm 0.01$}  
\newcommand{\srhoisochrones}{$3.43 \pm 0.08$} 
\newcommand{\Lisochrones}{$0.139\pm 0.005$}  

\newcommand{\sradiusSED}{$0.67\pm 0.03$}  
\newcommand{\AvSED}{$0.02\pm 0.02$}                
\newcommand{\FbolSED}{$7.63 \pm 0.18 \times 10^{-10}$}         
\newcommand{\smassSEDlogg}{$0.61 \pm 0.08$}         

\newcommand{\logRHKilaria}{$-4.86 \pm 0.03$} 
\newcommand{\protempirical}{$34 \pm 2$} 
\newcommand{\protspectroscopic}{$22 \pm 10$} 

\newcommand{\smasstorres}[1][]{$0.61\pm 0.04$}  
\newcommand{\sradiustorres}[1][]{$0.65\pm 0.05$}  
\newcommand{\sdensitytorres}[1][]{$3.07\pm 0.68$}  

\newcommand{\smassouthw}[1][]{$0.61\pm 0.05$}      
\newcommand{\sradiussouthw}[1][]{$0.63\pm 0.06$}   
\newcommand{\sdensitysouthw}[1][]{$3.40\pm 1.06$}  

\newcommand{\smasschweitzer}[1][]{$0.67\pm 0.02$}  

\newcommand{\smassboyajian}[1][]{$0.65\pm 0.04$}  

\newcommand{\sradiusparam}[1][]{$0.63\pm 0.02$}  
\newcommand{\smassparam}[1][]{$0.63\pm 0.02$} 
\newcommand{\sloggparam}[1][]{$4.64\pm 0.02$}     
\newcommand{\sdensityparam}[1][]{$3.53\pm 0.32$}     
\newcommand{\sAvparam}[1][]{$0.05^{+0.20}_{-0.18}$}     
\newcommand{\ageparam}{$8.4^{+4.7}_{-3.7}$}        

\newcommand{\citla}{\texttt{citlalicue}}
\newcommand{\george}{\texttt{george}}
\newcommand{\pyt}{\texttt{pytransit}}

\newcommand{\pyan}{\texttt{pyaneti}}

\newcommand{\rhostellar}[1][${\rm g^{1/3}\,cm^{-1}}$]{$3.47 _{ - 1.22 } ^ { + 0.89 }$~#1}
\newcommand{\Tzerob}[1][days]   {$8684.0128 _{ - 0.0024 } ^ { + 0.0016 }$~#1} 
\newcommand{\Pb}[1][days]   {$3.12748 _{ - 0.000038 } ^ { + 0.000047 }$~#1} 
\newcommand{\mpb}[1][$M_{\oplus}$]   {$8.6 _{ - 1.5 } ^ { + 1.4 }$~#1} 
\newcommand{\rpb}[1][$R_{\oplus}$]   {$2.34 _{ - 0.09 } ^ { + 0.11 }$~#1} 
  
\newcommand{\rrb}[1][ ]   {$0.0329 _{ - 0.0012 } ^ { + 0.0014 }$~#1} 
\newcommand{\arb}[1][ ]   {$12.14 _{ - 1.2 } ^ { + 0.7 }$~#1} 
\newcommand{\ab}[1][AU]   {$0.0366 _{ - 0.0036 } ^ { + 0.0022 }$~#1} 
\newcommand{\bb}[1][ ]   {$0.26 _{ - 0.17 } ^ { + 0.25 }$~#1}  

\newcommand{\kb}[1][${\rm m\,s^{-1}}$]   {$4.91 _{ - 0.83 } ^ { + 0.77 }$~#1} 
\newcommand{\denpb}[1][${\rm g\,cm^{-3}}$]   {$3.69 _{ - 0.76 } ^ { + 0.81 }$~#1} 
\newcommand{\grapb}[1][${\rm cm\,s^{-2}}$]   {$1520 _{ - 420 } ^ { + 370 }$~#1} 
\newcommand{\grapparsb}[1][${\rm cm\,s^{-2}}$]   {$1540 \pm 290$~#1} 
\newcommand{\Teqb}[1][K]   {$860 _{ - 31 } ^ { + 47 }$~#1} 
\newcommand{\insolationb}[1][${\rm F_{\oplus}}$]   {$91 _{ - 12 } ^ { + 22 }$~#1} 
\newcommand{\ib}[1][deg]   {$88.8 _{ - 1.4 } ^ { + 0.8 }$~#1}  
\newcommand{\ttotb}[1][hours]   {$1.963 _{ - 0.091 } ^ { + 0.066 }$~#1} 
\newcommand{\Tzeroc}[1][days]   {$8686.1179 _{ - 0.0035 } ^ { + 0.0033 }$~#1} 
\newcommand{\Pc}[1][days]   {$7.49325 _{ - 0.00013 } ^ { + 0.00015 }$~#1} 
\newcommand{\bc}[1][ ]   {$0.714 _{ - 0.066 } ^ { + 0.067 }$~#1} 
\newcommand{\kc}[1][${\rm m\,s^{-1}}$]   {$5.1 \pm 1.4$~#1} 
\newcommand{\mpc}[1][$M_{\oplus}$]   {$11.8 _{ - 3.2 } ^ { + 3.4 }$~#1} 
\newcommand{\rpc}[1][$R_{\oplus}$]   {$2.82 \pm 0.15 $~#1} 
\newcommand{\rrc}[1][ ]   {$0.0398 \pm 0.0020 $~#1} 
\newcommand{\ic}[1][deg]   {$88.12 _{ - 0.39 } ^ { + 0.24 }$~#1} 
\newcommand{\ttotc}[1][hours]   {$1.96 _{ - 0.10 } ^ { + 0.12 }$~#1} 
\newcommand{\arc}[1][ ]   {$21.7 _{ - 2.2 } ^ { + 1.2 }$~#1} 
\newcommand{\ac}[1][AU]   {$0.0656 _{ - 0.0065 } ^ { + 0.0039 }$~#1} 
  
\newcommand{\denstrc}[1][${\rm g\,cm^{-3}}$]   {$3.46 _{ - 0.93 } ^ { + 0.62 }$~#1} 
\newcommand{\insolationc}[1][${\rm F_{\oplus}}$]   {$28.4 _{ - 3.9 } ^ { + 6.8 }$~#1}  
\newcommand{\Teqc}[1][K]   {$643 _{ - 23 } ^ { + 35 }$~#1} 
\newcommand{\denpc}[1][${\rm g\,cm^{-3}}$]   {$2.87 _{ - 0.86 } ^ { + 0.98 }$~#1} 
\newcommand{\grapc}[1][${\rm cm\,s^{-2}}$]   {$1410 _{ - 500 } ^ { + 550 }$~#1} 
\newcommand{\grapparsc}[1][${\rm cm\,s^{-2}}$]   {$1450_{ - 410} ^ { + 450 }$~#1} 
\newcommand{\jPGP}[1][]   {$32.5 _{ - 2.2 } ^ { + 3.7 }$~#1}  
\newcommand{\jle}[1][]   {$45 _{ - 16 } ^ { + 17 }$~#1} 
\newcommand{\jlp}[1][]   {$1.4 _{ - 0.5 } ^ { + 1.0 }$~#1}  
\newcommand{\jArvc}[1][]   {$0.005 _{ - 0.004 } ^ { + 0.012 }$~#1}  
\newcommand{\jArvr}[1][]   {$0.22 _{ - 0.12 } ^ { + 0.32 }$~#1}  
\newcommand{\jAsmwc}[1][]   {$0.26 _{ - 0.12 } ^ { + 0.28 }$~#1} 
\newcommand{\qone}[1][]   {$0.44 _{ - 0.24 } ^ { + 0.33 }$~#1} 
\newcommand{\qtwo}[1][]   {$0.36 _{ - 0.24 } ^ { + 0.31 }$~#1} 
 
\newcommand{\qoneLCO}[1][]   {$0.35 _{ - 0.24 } ^ { + 0.39 }$~#1} 
\newcommand{\qtwoLCO}[1][]   {$0.42 _{ - 0.28 } ^ { + 0.32 }$~#1}
 
\newcommand{\HARPSN}[1][${\rm km\,s^{-1}}$]   {$0.0046 _{ - 0.0057 } ^ { + 0.0050 }$~#1}  
\newcommand{\Smw}[1][${\rm km\,s^{-1}}$]    {$1.11 \pm 0.17 $~#1}
\newcommand{\jHARPSN}[1][${\rm m\,s^{-1}}$]   {$0.88 _{ - 0.61 } ^ { + 0.83 }$~#1} 
\newcommand{\jSmw}[1][${\rm m\,s^{-1}}$]   {$0.0431 _{ - 0.0070 } ^ { + 0.0088 }$~#1} 
\newcommand{\jtr}[1][]   {$752 \pm 27 $~#1}  
\newcommand{\jtrLCO}[1][] {$141 _{ - 99 } ^ { + 15 }$~#1}

\newcommand{\rpd}[1][$R_{\oplus}$]   {$2.67 _{ - 0.25 } ^ { + 0.29 }$~#1} 
\newcommand{\Tzerod}[1][days]   {$1879.3211 _{ - 0.0055 } ^ { + 0.0067 }$~#1} 
\newcommand{\depthd}[1][ppm]   {$1418 _{ - 248 } ^ { + 317 }$~#1}

\newcommand{\vmic}{$V_{\rm mic}$}
\newcommand{\vmac}{$V_{\rm mac}$}

\newcommand{\nah}{[Na/H]}
\newcommand{\cah}{[Ca/H]}

\newcommand{\halpha}{H$\alpha$}                   

\newcommand{\lyalpha}{Ly$\alpha$}

\newcommand{\kms}{km\,s$^{-1}$}
\newcommand{\ms}{m~s$^{-1}$}

\newcommand{\gc}{g~cm$^{-3}$}
\newcommand{\lgr}{$\log\,(R^\prime_{HK})$}  

\title[TOI-1260]{Hot planets around cool stars – two short-period mini-Neptunes transiting the late K-dwarf TOI-1260}

\author[I.~Y.~Georgieva et al.]{I.~Y.~Georgieva,$^{1}$\thanks{E-mail: iskra.georgieva@chalmers.se}
C.~M.~Persson,$^{1}$,
O.~Barrag\'an$^{2}$,
G.~Nowak$^{3,4}$,
M.~Fridlund$^{1,5}$,
D.~Locci$^{6}$,
\newauthor
E.~Palle$^{3,4}$,
R.~Luque$^{3,4}$,
I.~Carleo$^{7}$,
D.~Gandolfi$^{8}$,
S.~R.~Kane$^{9}$,
J.~Korth$^{10}$,
K.~G.~Stassun$^{11}$,
\newauthor
J.~Livingston$^{12}$,
E.~C.~Matthews$^{13,14}$,
K.~A.~Collins$^{15}$,
S.~B.~Howell$^{16}$,
L.~M.~Serrano$^{8}$,
\newauthor
S.~Albrecht$^{17,18}$,
A.~Bieryla$^{15}$,
C.~E.~Brasseur$^{19}$,
D.~Ciardi$^{20}$,
W.~D.~Cochran$^{21}$,
K.~D.~Colon$^{22}$,
\newauthor
I.~J.~M.~Crossfield$^{23}$,
Sz.~Csizmadia$^{24}$,
H.~J.~Deeg$^{3,4}$,
M.~Esposito$^{25}$,
E.~Furlan$^{26}$,
T.~Gan$^{27}$,
\newauthor
E.~Goffo$^{8}$,
E.~Gonzales$^{28}$,
S.~Grziwa$^{29}$,
E.~.W.~Guenther$^{25}$,
P.~Guerra$^{30}$,
T.~Hirano$^{31,32}$,
\newauthor
J.~M.~Jenkins$^{16}$,
E.~L.~N.~Jensen$^{33}$,
P.~Kab\'ath$^{34}$,
E.~Knudstrup$^{17,18}$,
K.~W.~F.~Lam$^{35}$,
\newauthor
D.~W.~Latham$^{15}$,
A.~M.~Levine$^{13}$,
R.~A.~Matson$^{36}$,
S.~McDermott$^{37}$,
H.~L.~M.~Osborne$^{38}$,
\newauthor
M.~Paegert$^{15}$,
S.~N~.Quinn$^{15}$,
S.~Redfield$^{7}$,
G.~R.~Ricker$^{13}$,
J.~E.~Schlieder$^{39}$,
N.~J.~Scott$^{16}$,
\newauthor
S.~Seager$^{13,40,41}$,
A.~M.~S.~Smith$^{24}$,
P.~Tenenbaum$^{16,42}$,
J.~D.~Twicken$^{16,42}$,
R.~Vanderspek$^{13}$,
\newauthor
V.~Van~Eylen$^{38}$,
J.~N.~Winn$^{43}$
\newauthor\\
\\
Authors' affiliations are shown at the end of the manuscript}

\date{Accepted XXX. Received YYY; in original form ZZZ}

\pubyear{2020}

\begin{document}
\label{firstpage}
\pagerange{\pageref{firstpage}--\pageref{lastpage}}
\maketitle

\begin{abstract}
We present the discovery and characterization of two sub-Neptunes in close orbits, as well as a tentative outer planet of a similar size, orbiting \target{} -- a low metallicity \stype~dwarf star. Photometry from TESS yields radii of $R_{\rm b} = 2.33 \pm 0.10$ \rearth{} and $R_{\rm c} = 2.82 \pm 0.15$ \rearth, and periods of 3.13 and 7.49 days for \planetb{} and \planetc, respectively.
We combined the TESS data with a series of ground-based follow-up observations to characterize the planetary system. From HARPS-N high-precision radial velocities we obtain $M_{\rm b} = {} $\mpb{} and $M_{\rm c} = {} $\mpc. 
The star is moderately active with a complex activity pattern, which necessitated the use of Gaussian process regression for both the light curve detrending and the radial velocity modelling, in the latter case guided by suitable activity indicators. We successfully disentangle the stellar-induced signal from the planetary signals, underlining the importance and usefulness of the Gaussian Process approach.
We test the system's stability against atmospheric photoevaporation and find that the \target\ planets are classic examples of the structure and composition ambiguity typical for the $2-3$~\rearth\ range.

\end{abstract}

\begin{keywords}
Planetary systems --- planets and satellites: individual: \target b, c -- planets and satellites: atmospheres -- planets and satellites: composition -- techniques: photometric -- techniques: radial velocities, stars: low-mass

\end{keywords}



\section{Introduction}

Thanks to space-based photometry from missions like Convection, Rotation and planetary Transits \citep[CoRoT,][]{Baglin2006}, Kepler and K2 \citep{2010Sci...327..977B,2014PASP..126..398H} and Transiting Exoplanet Survey Satellite \citep[TESS,][]{2015JATIS...1a4003R}, the detection of shallow transits caused by small planets ($\lesssim 4$~\rearth) around faint stars has been made possible. The current exoplanet census shows that the most commonly detected population of planets is well represented by the so-called sub-Neptunes ($2 \lesssim R_\oplus \lesssim 4$) and rocky super-Earths ($1 \lesssim R_\oplus \lesssim 1.5$), with the radius valley \citep{2013ApJ...776....2L,2013ApJ...775..105O,2017AJ....154..109F,VanEylen2018,VanEylen21}, characterized by a paucity of planets between 1.5 and 2 \rearth\ \citep{2017AJ....154..109F}. This range has been shown to shift to smaller radii for low-mass stars \citep{FulPet18, Wu19, Cloutier2020, VanEylen21}.
An interesting observation about this population is the apparent ambiguity of the members' structures and compositions. \citet{Valencia07} first discussed the continuous wide range of planet compositions for a given mass and radius, while discrete reference planet models by \citet{Zeng2016,Zeng2019} show possible combinations of a rocky core with a H-He envelope, water-dominated worlds, as well as combinations of rock and ice bounded by H-He envelopes. This ambiguity is the result of the observed overlap between both the masses and radii of the two populations.
\citet{Otegi2020} report the transition range between sub-Neptunes to super-Earths to be $5-25$~\mearth\ and $2-3$~\rearth, which the \target\ planets presented in this work comfortably fall in.

Moving toward solving the aforementioned composition ambiguity would require understanding the dependence of close-in ($P_\mathrm{orb} < 10$~days) small ($2- 3$~\rearth) planets on parameters like the stellar mass \citep{FulPet18}, metallicity \citep{Wilson2018, Dong2018}, age \citep{Berger2020}, high-energy irradiation \citep{McDonald2019}, as well as the widely studied planetary mass, radius, period/semi-major axis. That said, while relatively precise radii are available from TESS, to place planets in the context of structure and composition models, we need precise mass estimates, and lots of them, as they are an indispensable piece of this puzzle.

The acquisition of precise masses is made possible thanks to high precision radial velocity (RV) measurements, performed by second generation spectrographs, such as ESO's HARPS \citep{Mayor03} and HARPS-N \citep{2012SPIE.8446E..1VC}, HIRES \citep{Vogt94}, CARMENES \citep{Quir14, Quir18}, and more recently ESPRESSO \citep{Pepe10, Pepe21}, EXPRES \citep{Jurg16} and more. Unfortunately, stellar activity can often be a complicating factor in obtaining accurate orbital solutions for the planet candidates. Great care and caution must be taken in accounting for this activity, the complexity of which may necessitate the use of more sophisticated methods than sinusoid fitting. This problem is further exacerbated the less massive and farther out from its star a planet is, as the precision required for a solid detection grows accordingly.


In this context, we present the discovery and characterization of the \target{} system -- a moderately active \stype~dwarf hosting two close-in ($P < 10$~days) transiting sub-Neptunes, as well as a tentative outer planet of similar size and an implied longer period. 

The paper is organized as follows. Section~\ref{sec:observations} contains a summary of the space and ground-based observations of \target\ as well as frequency analysis of the RVs and activity indicators, Sect.~\ref{sec:stellar} describes the stellar modelling, and in Sect.~\ref{sec:joint} we present our joint RV and transit analysis. In Sect.~\ref{sec:discussion} we discuss our findings and results and we summarize our conclusions in Sect.~\ref{sec:conclusions}.

\section{Observations}
\label{sec:observations}

Apart from space-based photometry from TESS, we obtained ground-based follow-up photometry from the Las Cumbres Observatory Global Telescope \citep[LCOGT,][]{Brown:2013}. We searched for stellar companions using Adaptive Optics (AO) and speckle imaging. To measure the planetary masses we observed \target{} with HARPS-N.

\subsection{TESS photometry}
\label{sec:photometry}

\begin{figure}
\centering
	\includegraphics[width=1\linewidth]{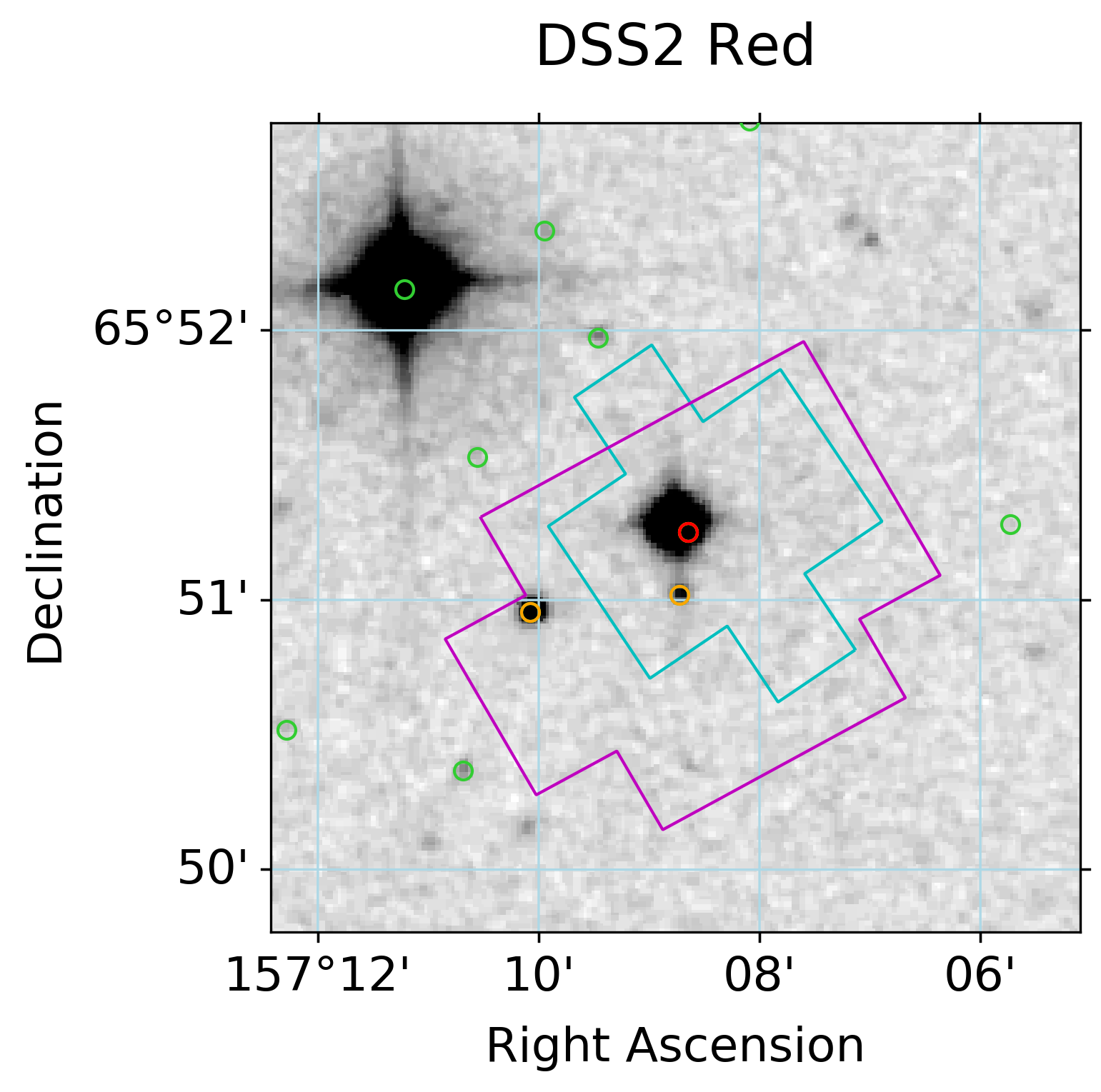}
    \caption{3\arcmin\,$\times$\,3\arcmin\ DSS2 (red filter) image with the Sectors 14 and 21 SPOC photometric apertures outlined in cyan and magenta, respectively. Colored circles denote the positions of \gaia\ DR2 sources within 2\arcmin\ of \target.}
    \label{fig:apts}
\end{figure}

\begin{figure*}
\begin{subfigure}{1\textwidth}
  \centering
  \includegraphics[width=1\linewidth]{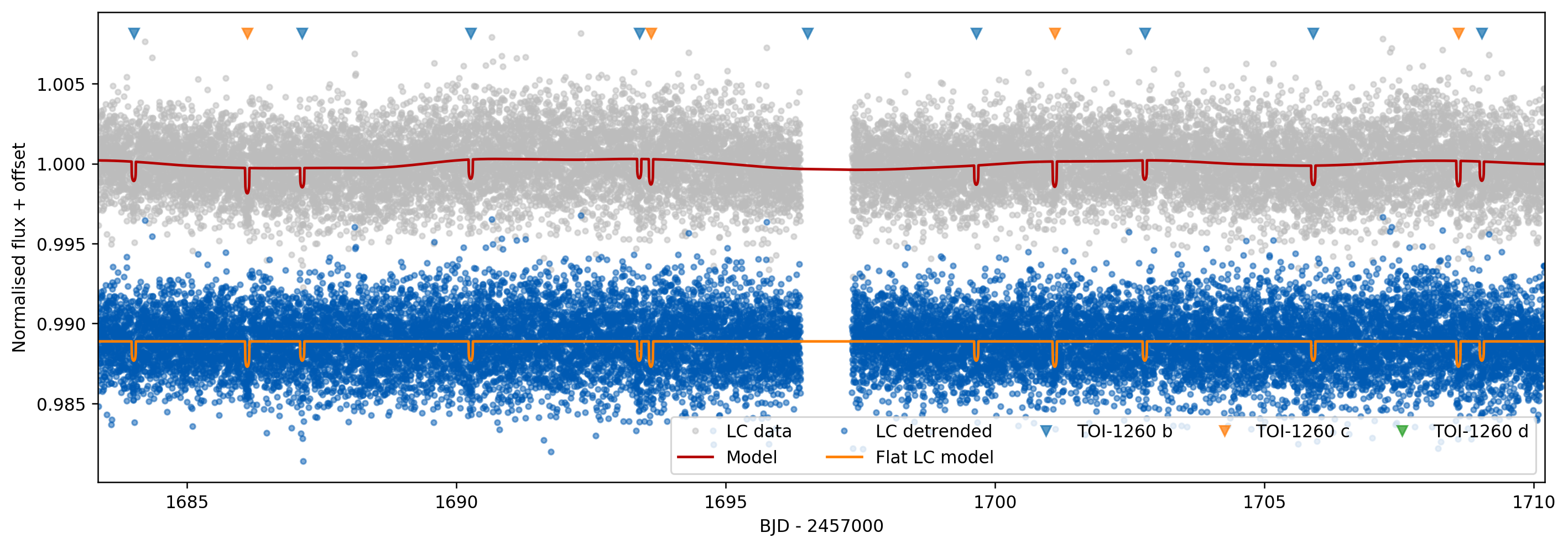} 
  \label{fig:s14}
\end{subfigure}
\vfill
\begin{subfigure}{1\textwidth}
  \centering
  \includegraphics[width=\textwidth]{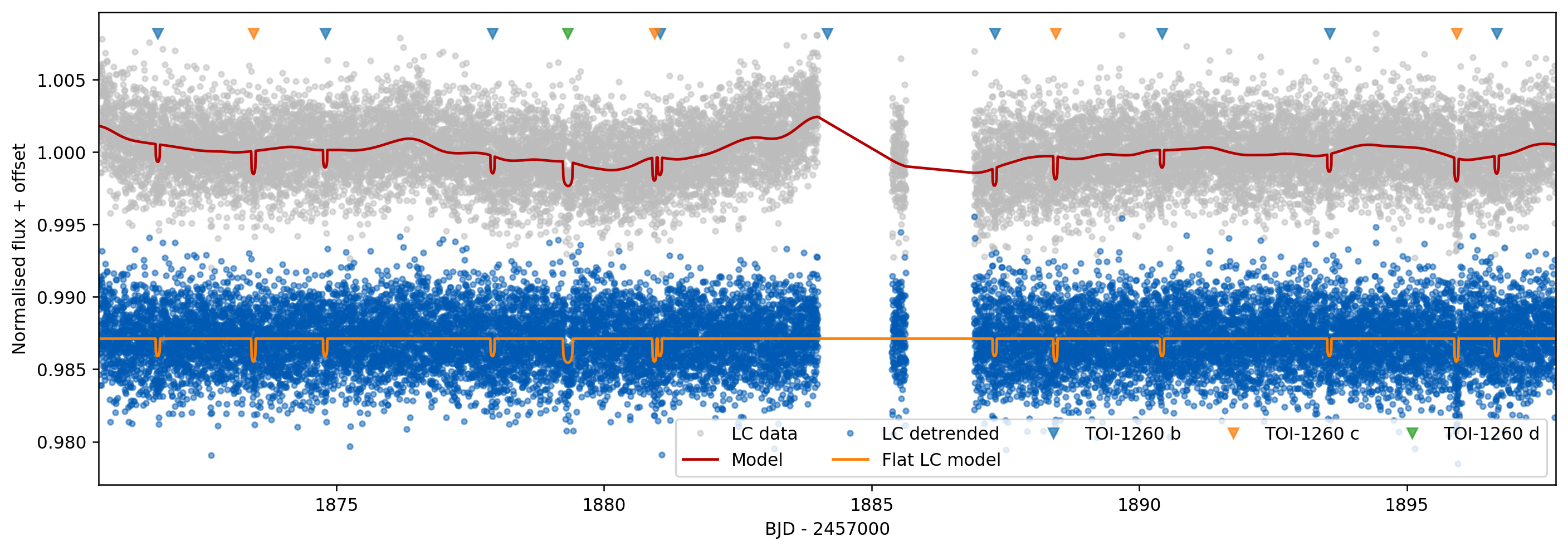}  
  \label{fig:s21}
\end{subfigure}
\caption{PDCSAP light curve in grey with GP model and transits overplotted in red, and resulting detrended light curve in blue for Sector 14 (top panel) and Sector 21 (bottom panel). The single transit event is visible in the bottom panel at 1879.3 days here plotted with a duration consistent with an arbitrary period of 40 days for visualization. Individual transits are marked with triangles.}
\label{fig:lc_all}
\end{figure*}

TESS first observed \target{} in Sector 14 between 2019 Jul 18 and 2019 Aug 15 on camera 4, CCD 3, and again in Sector 21 from 2020 Jan 21 to 2020 Feb 18 on camera 2, CCD 2. The target identifiers, coordinates, proper motion and magnitudes are listed in Table~\ref{tab:stellar}. Figure~\ref{fig:apts} shows a 3\arcmin$\,\times\,$3\arcmin\ digitized sky survey 2 (DSS-2, red filter) image centred on \target, marked by the red circle. The orange circles inside the Science Processing  Operations  Center \citep[SPOC, ][]{jenkins2016} apertures of the two sectors are potentially contaminating sources (TIC 841176092 with $V_{\rm mag}\approx19$ and TIC 138477027 with $V_{\rm mag}\approx16.2$ at 13.9\arcsec\ and 40\arcsec\ away from \target, respectively). However, the difference image centroid analyses performed for both TOIs detected in the SPOC pipeline, together with the ground-based follow-up observations discussed in the following sections, exclude this from being the case.
The SPOC pipeline \citep{Twicken10,Morris17} uses Simple Aperture Photometry (SAP) to generate stellar light curves, where common instrumental systematics, including dilution, are removed via the Presearch Data Conditioning (PDCSAP) algorithm \citep{Smith2012,Stumpe2012}. The TESS data were sampled at 2-min cadence and, after removing cadences flagged as potentially affected by anomalous events, the PDCSAP flux extracted from the FITS files produced by the SPOC pipeline (grey-dotted light curves in both panels of Fig.~\ref{fig:lc_all}) was used for both datasets to conduct the transit search.

Our transit search was realized via the MATLAB-based package EXOTRANS \citep{grziwa2012}. 
EXOTRANS utilizes filtering routines based on the Stationary Wavelet Transform to remove intrinsic stellar variability as well as signals at known frequencies to allow searching for additional transits. 
The search itself is performed using an optimized version of the traditional well-established BLS algorithm \citep{kovacs2002}, as described in \citet{Ofir2014}.
TOIs 1260.01 and 1260.02 were first discovered in the SPOC transit search \citep{Jenkins02,Jenkins10,Jenkins17} with periods of 3.13 and 7.49~days, respectively, and announced in the TESS SPOC data validation reports \citep[DVR,][]{Twicken2018} and the TOI release portal\footnote{https://tess.mit.edu/toi-releases/}. We note that 1260.02 is missing from the DVR for Sector 21. Instead, in addition to 1260.01, a signal at 16.613 days was reported but was not given TOI status, likely due to the significant difference in depth between its two apparent transits, the second of which coincides with a transit of 1260.02. This is further discussed in Sect.~\ref{sec:Planetd}.

EXOTRANS detected the two candidates with depths of 1222 ppm and 1685 ppm in both TESS sectors, and periods in agreement with the publicly announced 1260.01 and 1260.02, respectively. 
As an additional check, we further analysed the light curve data using the {\tt lightkurve} package \citep{lightkurve}. We discovered no significant odd/even difference or a sign of a secondary eclipse. This concurs with the results in the DVRs, where the odd/even depth test and difference image centroid test also found no evidence for either signal being due to an eclipsing binary or background eclipsing binary. Encouraged by the agreement between the different pipelines, we prioritized \target\ and qualified it as a promising target for follow-up observations.

Due to the complex variability \target\ exhibits, we chose to remove the low frequency signals in the light curves using a Gaussian process (GP). We use the Python package  \href{https://github.com/oscaribv/citlalicue}{\citla}\footnote{https://github.com/oscaribv/citlalicue}, which is a wrapper of \href{https://github.com/dfm/george}\george\ \citep{Mackey2014, Ambi2016} and  \href{https://github.com/hpparvi/PyTransit}\pyt\ \citep{Parviainen2015}. Briefly, \citla\ performs a GP regression (given a covariance function as provided by {\tt{george}}) together with transit models ({\tt{\pyt}}) to the data. The best fitting model is computed by likelihood maximization. This generates a model that contains variability and transits. \citla\ then removes the light curve variability model from the data to create a flattened normalized light curve with only transits.

We ran \citla\ with a GP created with a Mat\'ern 3/2 covariance function together with a model of the two transiting planet candidates and an additional single transit we identified in Sector 21 at T$_0$~$\sim1879.32$. Since we are not interested in the nature of the variability signal, we chose the Mat\'ern 3/2 kernel because of its flexibility in dealing with stochastic correlation. We performed individual runs for each sector given that light curve variability scales may be different between the sectors.

The PDCSAP light curves of both sectors are shown in Fig.~\ref{fig:lc_all}, along with the flattened light curves and transit models. We use these flattened light curves for our joint analysis in Sect.~\ref{sec:joint}. The single transit is visible in the lower panel of Fig.~\ref{fig:lc_all} and its depth is approximately 1430 ppm. The feature is shown plotted assuming an arbitrary period of 40 days, which is within the range of possible periods for this possible outer planet (more on this in Sect.~\ref{sec:Planetd}).

\begin{table}
\centering
\caption{Main identifiers, equatorial coordinates, proper motion, parallax, optical and infrared magnitudes, and fundamental parameters of \target.}
\label{tab:stellar}
\begin{tabular}{lrr}
\hline
Parameter & Value & Source \\
\hline
\multicolumn{3}{l}{\it Main identifiers}  \\
\noalign{\smallskip}
\multicolumn{2}{l}  {TIC} {355867695}  & ExoFOP$^a$ \\
\multicolumn{2}{l}{2MASS}{J10283500+6551163}  & ExoFOP \\
\multicolumn{2}{l}{UCAC4} {780-023265}  & ExoFOP\\
\multicolumn{2}{l}{WISE} {J102834.71+655115.5}  & ExoFOP\\
\multicolumn{2}{l}{APASS} {59325479}  & ExoFOP\\
\hline
\multicolumn{3}{l}{\it Equatorial coordinates, parallax, and proper motion}  \\
\noalign{\smallskip}
R.A. (J2000.0)	&    10$^\mathrm{h}$28$^\mathrm{m}$34.56$^\mathrm{s}$	& {\it Gaia} DR3$^b$ \\
Dec. (J2000.0)	& $+$65$\degr$51$\arcmin$15.07$\arcsec$	                & {\it Gaia} DR3 \\
$\pi$ (mas) 	& $13.6226\pm0.0147$                                    & {\it Gaia} DR3 \\
$\mu_\alpha$ (mas\,yr$^{-1}$) 	& $-177.340 \pm 0.012$		& {\it Gaia} DR3 \\
$\mu_\delta$ (mas\,yr$^{-1}$) 	& $-81.693 \pm 0.013$		& {\it Gaia} DR3 \\
\hline
\multicolumn{3}{l}{\it Optical and near-infrared photometry} \\
\noalign{\smallskip}
$TESS$              & $10.812\pm0.006$     & TIC v8$^c$         \\
\noalign{\smallskip}
$G$				 & $11.5655\pm0.0.0028$ $^d$	& {\it Gaia} DR3 \\
$B_\mathrm{p}$   & $12.2955\pm0.0030$ $^d$  & {\it Gaia} DR3 \\
$R_\mathrm{p}$   & $10.7415\pm 0.0038$ $^d$  & {\it Gaia} DR3 \\
\noalign{\smallskip}
$B$              & $13.259 \pm 0.088$          & APASS \\
$V$              & $11.875 \pm 0.165$          & APASS \\
$g$              & $12.702 \pm 0.060$          & APASS \\
\noalign{\smallskip}
$J$ 			&  $9.698\pm0.023$      & 2MASS \\
$H$				&  $9.105\pm0.027$      & 2MASS \\
$Ks$			&  $8.950\pm0.022$      & 2MASS \\
\noalign{\smallskip}
$W1$			&  $8.891\pm0.023$      & All{\it WISE} \\
$W2$			&  $8.964\pm0.020$      & All{\it WISE} \\
$W3$             & $8.880\pm0.023$      & All{\it WISE} \\
$W4$             & $9.215\pm0.453$      & All{\it WISE} \\
\hline
\multicolumn{3}{l}{\footnotesize$^a$\href{https://exofop.ipac.caltech.edu/}{https://exofop.ipac.caltech.edu/}} \\
\multicolumn{3}{l}{\footnotesize$^b$\citet{2021A&A...649A...6G}} \\
\multicolumn{3}{l}{\footnotesize$^c$\citet{Stassun18}} \\
\multicolumn{3}{l}{\footnotesize$^d$Uncertainties from the VizieR Catalogue, \citet{2000A&AS..143...23O}}.\\
\end{tabular}
\end{table}

\subsection{Light curve follow-up}
\label{sec:TFOP}
As a further step towards confirming the planets and to try and improve the system parameters, we acquired ground-based time-series follow-up photometry of \target\ as part of the TESS Follow-up Observing Program (TFOP)\footnote{https://tess.mit.edu/followup}. We used the {\tt TESS Transit Finder}, which is a customized version of the {\tt Tapir} software package \citep{Jensen:2013}, to schedule our transit observations. The photometric data were extracted using {\tt AstroImageJ} \citep{Collins:2017}.

\subsubsection{LCOGT}
We observed a full transit of 1260.01 on 2020 Jan 04 and parts of the 1260.02 SPOC ephemeris $3\sigma$ window on 2019 Dec 03 and 2020 February 01 from LCOGT 1.0\,m network node at McDonald Observatory. All observations were in the Pan-STARSS $z$-short filter. The $4096\times4096$ LCOGT SINISTRO cameras have an image scale of $0\farcs389$ per pixel, resulting in a $26\arcmin\times26\arcmin$ field of view. The 1260.01 images were defocused and have typical stellar point-spread-functions (PSFs) with full-width-half-maximum (FWHM) $\sim8\farcs3$, and circular apertures with radius $\sim 9\farcs 7$ were used to extract the differential photometry. 
Regarding both epochs of TOI 1260.02, the first observations cover a partial (half) transit, and on the second occasion the observations cover a fraction of the transit ingress. Neither dataset shows a hint of the planet signal. This can be caused by data reduction systematics given the partial coverage of the transits and the relatively low light curve precision. Therefore we do not use these data for further analysis.
The photometry ruled out a transit on target and ruled out possible contaminating nearby eclipsing binaries (NEBs) within $2\farcm5$ of the target star over the observing window.

\subsubsection{KeplerCam}
We observed overlapping transits of TOIs 1260.01 and 1260.02 (assuming the initial SPOC Sector 14 nominal ephemerides) in Sloan $i'$-band on 2019 November 18 from KeplerCam on the 1.2\,m telescope at the Fred Lawrence Whipple Observatory. The $4096\times4096$ Fairchild CCD 486 detector has an image scale of $0\farcs336$ per pixel, resulting in a $23\farcm1\times23\farcm1$ field of view. The observations were focused and the resulting images have typical stellar PSFs with a FWHM of $\sim1\farcs5$. Circular apertures with radius $\sim 4\farcs 7$ were used to extract the differential photometry. The on-target light curve was inconclusive, but possible contaminating NEBs within $2\farcm5$ of the target star were ruled out over the 183 minute observing window.

\subsection{AO with Gemini-North/NIRI}

\begin{figure}
\begin{subfigure}{0.5\textwidth}
\includegraphics[width=1\linewidth]{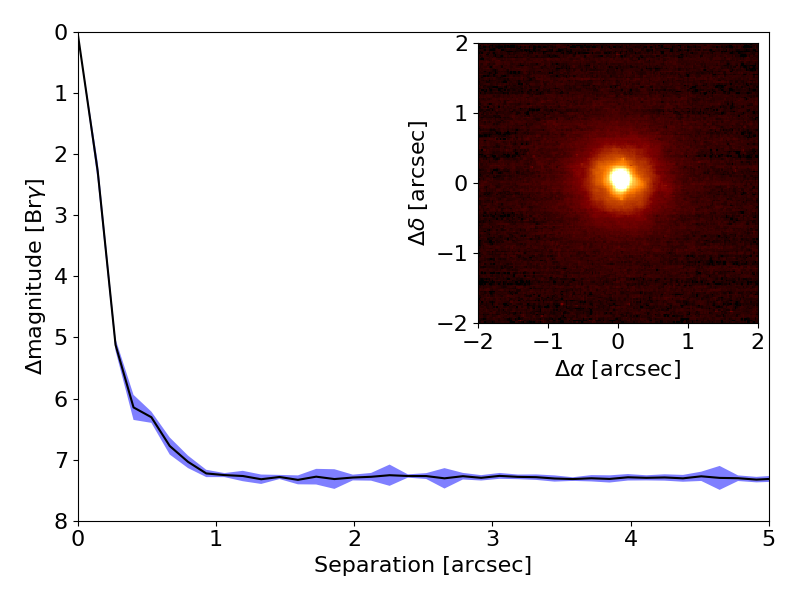}
\includegraphics[width=1\linewidth]{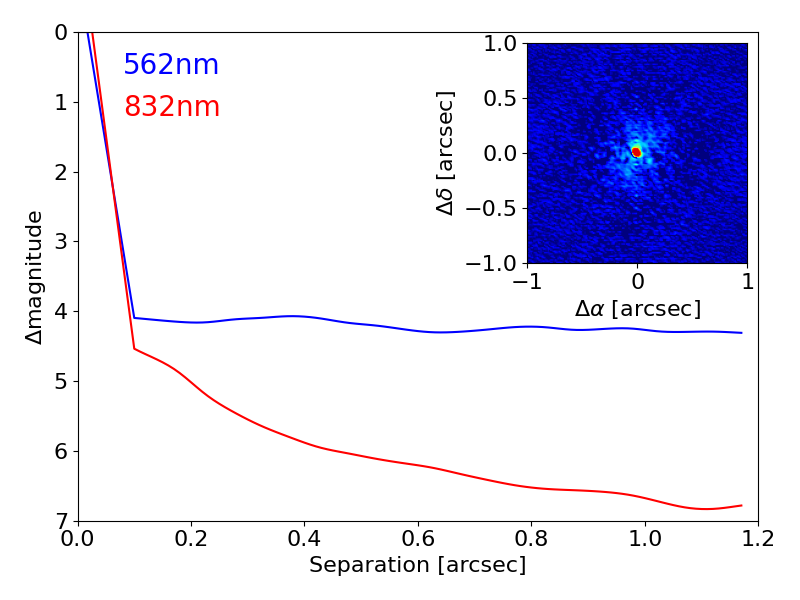}
\end{subfigure}
\caption{Upper panel: sensitivity to faint visual companions of our Gemini/NIRI observations of \target. Companions 5 magnitudes fainter than the host star can be detected beyond 270~mas and no companions are seen anywhere in the field of view, which extends at least 13\arcsec~from the target in all directions. The inset shows the central portion of the image, centered on the star, and the star appears single to the limit of our resolution. Lower panel: 5-$\sigma$ sensitivity curve of speckle imaging by Gemini North/‘Alopeke showing a reconstructed image of the field. No bright companions are detected within 1.2\arcsec.}
\label{fig:AOspeck}
\end{figure}

It is crucial that close visual companions are identified, since these can dilute the lightcurve and thus alter the planet properties, or even be the source of false positive signals, in the case that the visual companion is itself a binary \citep[see e.g.][]{ciardi2015}. We search for such companions using AO imaging using the NIRI instrument \citep{hodapp2003} at the Gemini-North telescope. We collected a total of 9 images of \target\ on 2019 Nov 25, using the narrow-band Br$\gamma$ filter which falls within the K-band. Each image had an exposure time of 3.9~s, and we dithered the telescope between each image. This allows for a sky background frame to be constructed from the science data itself, by median combining these dithered frames. Our data reduction process consisted of bad pixel removal, flat-correction and sky-background subtraction, and aligning the stellar position between frames so they could be coadded. We searched for companions in the final image visually, and did not identify companions anywhere in the field of view, which extends to at least 13\arcsec~from the star in all directions. 
We used a fake star injection technique to measure the sensitivity of the data. In this process we sequentially injected fake PSFs (constructed from the measured stellar PSF, and with peak brightness 3 times the local dispersion level) into the image, every 132~mas in the radial direction and at 8 distinct position angles for each radius. We measured the significance of each fake PSF, and linearly scale this value to the flux at which a companion would be detected with 5$\sigma$ significance. The quoted sensitivity at each radius is the median sensitivity across the 8 position angles. 
We are sensitive to companions 5 magnitudes fainter than the star at separations beyond 270~mas, and reach a contrast limit of $\Delta K = 7.3$~mag in the wide field. The upper panel in Fig.~\ref{fig:AOspeck} shows the sensitivity of our survey, and the inset shows an image of the target itself.

We note that the above described procedure has been used in a wide range of papers \citep[see e.g.][]{2019NatAs...3.1099G, 2019AJ....157..191R, 2019AJ....158...32K}.

\subsection{Gemini-North/‘Alopeke speckle imaging}

While AO imaging is sensitive in the infrared and at wider separations from the target, speckle imaging explores the closer vicinity of the target at optical wavelengths.

\target{} was observed on 2020 Feb 16 using the ‘Alopeke speckle instrument on Gemini-North\footnote{https://www.gemini.edu/sciops/instruments/alopeke-zorro/}. 
‘Alopeke provides simultaneously speckle imaging in two bands, 562\,nm and 832\,nm, with output data products including a reconstructed image, and robust limits on companion detections \citep{howell2011}.
Figure ~\ref{fig:AOspeck} (lower panel) shows our resulting contrast curves and  the reconstructed 832 nm speckle image. 
We find that \target{} is a single star with no companion brighter than about 5\,-\,7 magnitudes detected within 1\farcs2.  
‘Alopeke observations provide resulting spatial resolutions of 0.017~mas in the blue, and 0.026~mas in the red,  yielding an inner working angle of 1.18 and 1.84 au at the distance to \target, respectively.

\subsection{High-dispersion spectroscopy with TNG/HARPS-N}
\label{sec:RVobs}

Currently, RV measurements are invaluable for the purpose of planetary mass determination. Such observations, however, also allow for co-added stellar spectra to be obtained, which are used to model the star and thus obtain more accurate stellar parameters.

Between 2020 Jan 14 and 2020 June 13 we collected 33 spectra with the HARPS-N spectrograph \citep[R$\approx$115\,000]{2012SPIE.8446E..1VC} mounted at the 3.58-m Telescopio Nazionale Galileo (TNG) of Roque de los Muchachos Observatory in La Palma, Spain, under the observing programmes CAT19A\_162, ITP19\_1 and A40TAC\_22\footnote{20 spectra were obtained from the Spanish CAT19A\_162 programme (PI: Nowak), 12 spectra from ITP19\_1 programme (PI: Pall\'e) and one spectrum from A40TAC\_22 programme (PI: Gandolfi).}.
The exposure time was set to 1350\,--\,3600~s, based on weather conditions and scheduling constraints, leading to a SNR per pixel of 21\,--\,74 at 5500\,\AA. 
The spectra were extracted using the off-line version of the HARPS-N Data Reduction Software (DRS) pipeline \citep{2014SPIE.9147E..8CC}, version 3.7. 
Absolute RVs and spectral activity indicators -- bisector inverse slope (BIS), full-width at half maximum (CCF\_FHWM), contrast (CCF\_CTR) of the cross-correlation function (CCF) and Mount-Wilson S-index -- were measured using an on-line version of the DRS, the \href{http://ia2-harps.oats.inaf.it:8000}{YABI} tool, by cross-correlating the extracted spectra with a K5 mask \citep{1996A&AS..119..373B}. We also used \href{https://github.com/mzechmeister/serval}{\tt serval} \citep{2018A&A...609A..12Z} code to measure relative RVs by the template-matching, chromatic index (CRX), differential line width (dLW), and H$\alpha$ index. 
The uncertainties of the RVs measured with {\tt serval} are in the range 0.9\,--\,3.1\,\mps, with a mean value of 1.6\,\mps. 
Table~\ref{all_rv.tex} gives the time stamps of the spectra in BJD$_{\mathrm{TDB}}$, {\tt serval} relative RVs along with their $1\sigma$ error bars, and spectral activity indicators measured with YABI and  {\tt serval}. In the joint RV and transit analysis presented in Section 5 we used relative RVs measured from HARPS-N spectra with {\tt serval} by the template-matching technique.

\subsubsection{Frequency analysis of TNG/HARPS-N data}
\label{sec:freq}

\begin{figure}
\centering
\includegraphics[width=\linewidth, height=19.5cm]{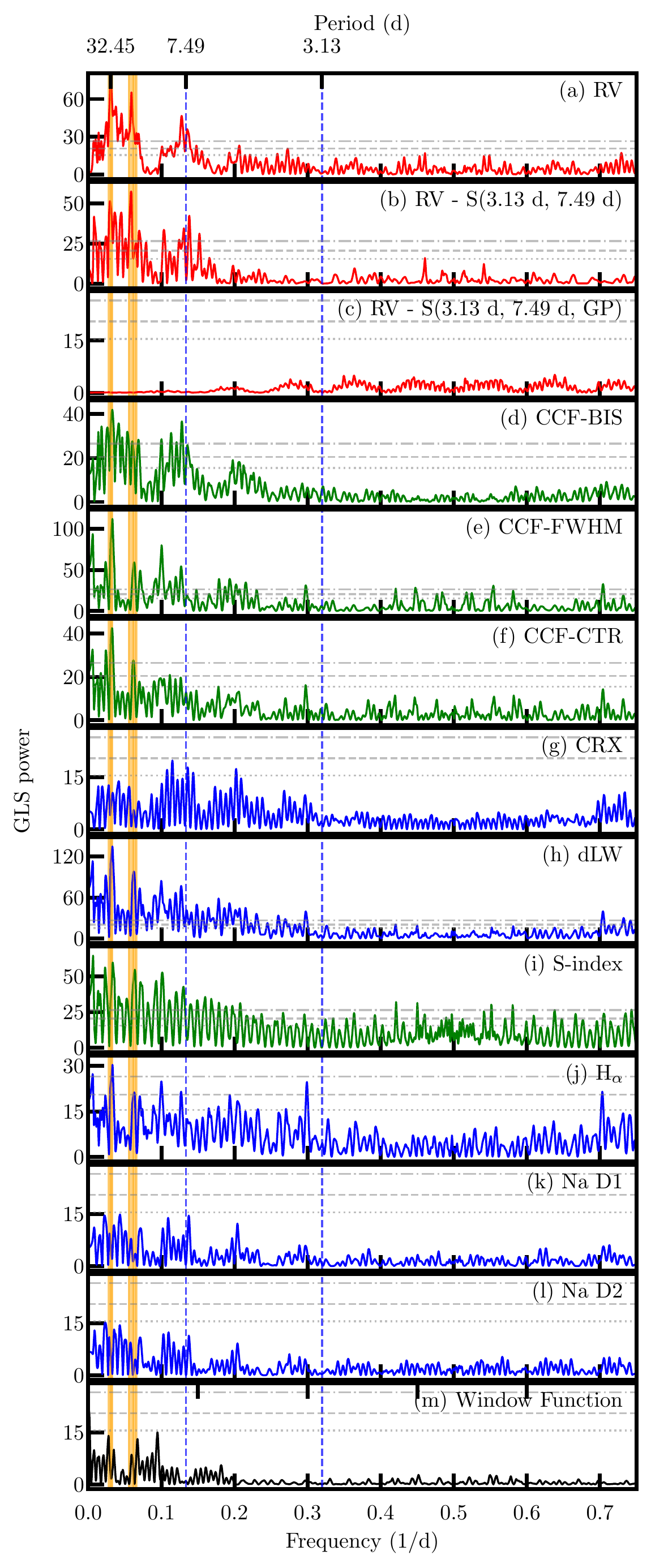}
    \caption{Generalized Lomb-Scargle periodograms of RVs of \target{} (a), their residuals (b) after fitting two sinusoids with periods and phases corresponding to 1260.01 ($f_\mathrm{b} = 0.320\pm0.002\,\mathrm{days}^{-1}$, $P_\mathrm{b} = 3.13\pm0.02\,\mathrm{days}$) and 1260.02 ($f_\mathrm{c} = 0.133\pm0.002\,\mathrm{days}^{-1}$, $P_\mathrm{c} = 7.49\pm0.11\,\mathrm{days}$), marked as vertical blue dashed lines, and their residuals (c) after fitting final joint model presented in Sect.~\ref{sec:joint}. Vertical orange areas present frequency of the GP signal ($f_\mathrm{GP} = 0.031^{+0.002}_{-0.003}\,\mathrm{days}^{-1}$, $P_\mathrm{GP} = 32.45^{+3.70}_{-2.14}\,\mathrm{days}$) and its first harmonic. Panels plotted in green show periodograms of spectral activity indicators measured with DRS pipeline and panels plotted in blue activity indicators measured with {\tt {serval}}. Last panel (m) presents the window function of the data. Horizontal grey lines show the theoretical FAP levels of 10\,\% (dotted line), 1\,\% (dashed line), and 0.1\,\% (dash-dotted line) for each panel.
\label{figure-TOI-1260-glsp}}
\end{figure}

In order to search for the Doppler reflex motion induced by the transiting planetary candidates and unveil the presence of possible additional signals we performed a frequency analysis of the RVs and spectral activity indicators measured from TNG/HARPS-N spectra. We calculated the generalised Lomb-Scargle (GLS) periodograms \citep{Zech09} of the available time series and computed the theoretical 10\,\%, 1\,\%, and 0.1\,\% false alarm probability (FAP) levels (Fig.~\ref{figure-TOI-1260-glsp}). The 151.8~day time baseline of the measurements translate into a frequency resolution of 0.006586~days$^{-1}$.

The strongest peak in the GLS periodogram of RVs \mbox{($\mathrm{FAP}<0.1\%$)} has a frequency of $\sim$0.031, i.e. a period of $\sim$32.5~days (panel (a) of Fig.~\ref{figure-TOI-1260-glsp}). Peaks at this frequency are also the strongest ones in the GLS periodograms of spectral activity indicators measured with the DRS pipeline, especially in the periodogram of CCF-FWHM (panel (e) of Fig.~\ref{figure-TOI-1260-glsp}) and in the periodogram of dLW measured with {\tt {serval}} (panel (h) of Fig.~\ref{figure-TOI-1260-glsp}). The GLS periodogram of residuals after fitting two sinusoids with periods and phases corresponding to 1260.01 ($f_\mathrm{b} = 0.320\pm0.002\,\mathrm{days}^{-1}$, $P_\mathrm{b} = 3.13\pm0.02~\mathrm{days}$) and 1260.02 ($f_\mathrm{c} = 0.133\pm0.002~\mathrm{days}^{-1}$, $P_\mathrm{c} = 7.49\pm0.11~\mathrm{days}$) shows two highly significant peaks (FAP\,<\,0.1\,\%) at the frequency of $0.031^{+0.002}_{-0.003}\mathrm{days}^{-1}$ and its first harmonic. This clearly shows that the strongest signal in the radial velocities has its origin in stellar activity. The RV residuals after a joint model presented in Sect.~\ref{sec:joint} (panel (c) of Fig.\ref{figure-TOI-1260-glsp}) show no further significant peaks. In the GLS periodograms of the activity indicators there are no peaks at the frequencies of the candidates.

The above results show that due to the suboptimal quantity and sampling of the data, a simple periodogram inspection is not suitable for such subtle and sophisticated analysis as required by this system. For the global model we thus implement a more advanced technique as demonstrated in Sect.~\ref{sec:joint}.

\section{Stellar Modelling}
\label{sec:stellar}
\subsection{Spectral analysis}
\label{Sect: Spectral analysis}
We modelled the co-added  high resolution ($R = 115\,000$) HARPS-N spectra with a signal-to-noise of 125 at 
5800~\AA~with the spectral analysis package \href{http://www.stsci.edu/~valenti/sme.html}{\tt{SME}}
\citep[Spectroscopy Made Easy;][]{vp96, pv2017} version 5.22. 
This software package matches observations to 
synthetic stellar spectra calculated from grids of atmosphere models  
using a $\chi^2$-minimising procedure.  We used the  {\tt{MARCS 2012}} \citep{Gustafsson08} grid  and
also checked the final models with the  {\tt{ATLAS12}} model spectra \citep{Kurucz2013}. 
The line data was taken from \href{http://vald.astro.uu.se}{VALD} \citep{Ryabchikova2015}. 
We derived  the effective temperature (\teff), the stellar surface gravity (\logg), abundances,  
the projected stellar rotational velocity (\vsini), 
and  the macroturbulent velocity (\vmac), following  the procedures described in \citet{2018A&A...618A..33P} and 
\citet{2017A&A...604A..16F}. In summary, we used   
the line wings of H$\alpha$  to derive \teff, and   
 \logg~was modelled with the line wings of the \ion{Ca}{I} 
$\lambda \lambda$6102, 6122, and 6162 triplet, and the $\lambda$6439 line.
Due to the low \teff, and hence the weak line wings of H$\alpha$ and the large number of metal lines contaminating 
the diagnostic line wings, we also used the Na doublet 
$\lambda \lambda$5889 and 5896 sensitive to both \teff~and \logg~to check our model. 
\vsini,   \vmac, and the iron and calcium abundances, 
\feh~and \cah, were modelled  with 
narrow and unblended lines between $\lambda$6000 and $\lambda$6500, and the \nah~abundance 
with lines  between $\lambda$5600 and $\lambda$6200. The abundances of   Ca and Na were similar to Fe.          
The macroturbulent and radial velocities were found to be \svmacsme~\kms~and 
-16.6~\kms, respectively, while the 
microturbulent velocity, \vmic, was fixed to 1~\kms. 

To check the {\tt{SME}} results we also used the empirical  {\tt{SpecMatch-Emp}}   \citep{2017ApJ...836...77Y} code 
characterising   stars based on their optical spectra. The software compares the observed spectrum to   
a   spectral library of  more than 400 well-characterised  stars with spectral classes M5 to F1 observed by Keck/HIRES.
Since the library stars often 
have their radii calibrated using interferometry, the direct output is \teff, \rstar, and \feh. 
Before running the code, we    transformed  our co-added HARPS-N  spectra into the  
format of Keck/HIRES spectra used by {\tt{SpecMatch-Emp}} as outlined in \citet{2018AJ....155..127H}.

The models are in excellent agreement and we list the results  in Table~\ref{tab:comparison stellar spectroscopic parameters} 
along with the effective temperature from Gaia as a comparison. We adopt the {\tt{SME}} results for the modelling of the stellar
mass and radius in the following section.

\subsection{Stellar mass and radius}
\label{Sect: Stellar mass and radius}
\begin{figure}
\centering
	\includegraphics[width=\linewidth]{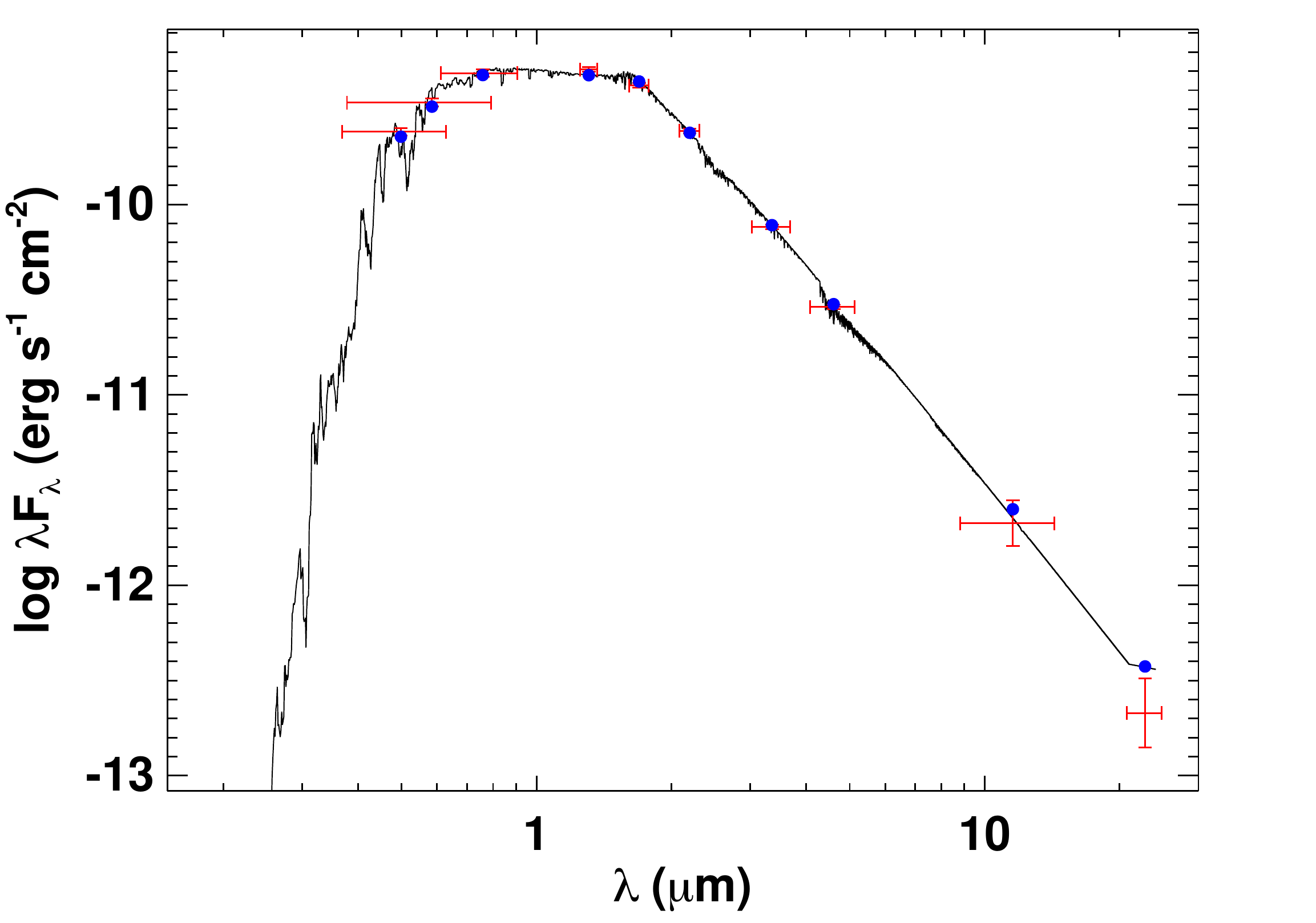}
    \caption{Spectral energy distribution of \target. Red symbols represent the observed photometric measurements, where the horizontal bars represent the effective width of the passband. Blue symbols are the model fluxes from the best-fit NextGen atmosphere model (black).}
    \label{fig:sed}
\end{figure}

We started with an independent determination of the   stellar radius, and performed an analysis of the broadband spectral 
energy distribution (SED) of the star together with the {\it Gaia\/} DR2 parallaxes 
adjusted by $+0.08$~mas to account for the systematic offset reported by 
\citet[][]{2018ApJ...862...61S}. 
We followed the procedures described in \citet{2016AJ....152..180S} and \citet{2017AJ....153..136S,2018AJ....155...22S} 
and pulled the $JHK_S$ magnitudes from the {\it 2MASS} catalogue, the $W1-W4$ magnitudes from the {\it WISE} catalogue, 
and the $G G_{\rm BP} G_{\rm RP}$ 
magnitudes from the {\it Gaia} database. Together, the available photometry spans the stellar 
SED over the wavelength range 0.4--22~$\mu$m.  
We performed a fit using NextGen stellar atmosphere models, with $T_{\rm eff}$, [Fe/H], and $\log g$ 
adopted from the spectroscopic analysis with {\tt{SME}} as priors. The only additional free
parameter is the extinction ($A_V$), which we restricted to the maximum line-of-sight
value from the dust maps of \citet{1998ApJ...500..525S}. The resulting fit, shown in
Fig.~\ref{fig:sed}, is very good with a reduced $\chi^2$ of 1.1 and best-fit 
$A_V$\,=\,\AvSED. Integrating the (unreddened) SED model gives the bolometric flux at
Earth, $F_{\rm bol}$\,=\,\FbolSED~erg~s$^{-1}$~cm$^{-2}$. Taking the $F_{\rm bol}$ 
and $T_{\rm eff}$ together with the {\it Gaia\/} DR2 parallax, gives the stellar
radius. Using this radius  together with the
spectroscopic $\log g$, we obtain an empirical mass estimate. 

In order to obtain  a uniform set of stellar parameters we used the Python code
\href{https://github.com/timothydmorton/isochrones}{\tt{isochrones}} \citep{2015ascl.soft03010M}, an   
MCMC fitting tool of stellar properties based on
an interface interacting with  the MIST \citep{2016ApJ...823..102C}  stellar evolution tracks. 
We fitted the $Gaia$ DR2 parallax and   the 2MASS $JHK$ photometry, the four WISE magnitudes and the $B$- and $V$-bands 
from APASS, with priors on \teff, \logg, and \feh~from
{\tt{SME}} using {\tt{MultiNest}} \citep{2014A&A...564A.125B}\ to sample the joint posteriors. We find a bolometric luminosity
of \Lisochrones~\lsun. 
 
The above results were checked with the Bayesian 
\href{http://stev.oapd.inaf.it/cgi-bin/param}{\tt{Param\,1.5}} \citep{daSilva2006} on-line code 
using the  {\tt{PARSEC}} isochrones \citep{2012MNRAS.427..127B} and the same input as for {\tt{isochrones}}.  
 
We also computed mass and radius from the empirical calibration 
equations by \citet{2010AJ....140.1158T} from \teff, \logg, and \feh. 
Finally, we used the  stellar mass-radius relations  for 
low-mass stars from 
\citet{2012ApJ...757..112B}   to compute the stellar mass from the radius obtained from  {\tt{isochrones}}.
  
The stellar parameters found above indicate that this star is a \stype~star supported by  the 
empirical relations of \citet{2012ApJ...756...47S} suggesting that the activity-driven radius 
inflation is at most $\sim$2\%, indicating a star on the main-sequence. This is also consistent with the age estimates with 
{\tt{Param\,1.5}} of 
\ageparam~Gyr.

All results of the stellar mass and radius are in very  good agreement and are listed in Table~\ref{tab:comparison stellar parameters} along with
a typical mass and radius for an \stype~star for comparison. 
We adopt the stellar mass and radius from  {\tt{isochrones}} in our joint modelling of the  
system in Sect.~\ref{sec:joint} and list our adopted parameters for the modelling in Table~\ref{params}.

\subsection{Stellar activity and rotation period}
\label{Sect: actandrot}
We note that both  \ion{Ca}{II} H $\&$ K lines are seen in emission  in the HARPS-N spectra which indicates 
that the star is moderately active. The activity offers a way to estimate  the rotation period. 
We first computed the average \smw~from the time series to be \mbox{$1.13 \pm 0.08$} which was converted to
\lgr\,=\,\logRHKilaria~\citep{2015MNRAS.452.2745S}. 
This was used together with the empirical relation for late-type stars from \citet{2015MNRAS.452.2745S, 2016MNRAS.457.2604S} 
and the star's color to predict a rotation period of \protempirical~days.
This is  within 1~$\sigma$ of the \mbox{\protspectroscopic~days} estimate obtained from \rstar~together with the spectroscopically determined \vsini,
assuming that the star is equator-on oriented. 

The   activity predicts an age of $4.1\pm0.2$~Gyr, from the empirical relations of 
\citet{2008ApJ...687.1264M} which is considerably lower than derived above although still within the large uncertainties. 
The estimate from gyrochronology has, however, the following two caveats: this star is 
somewhat cooler than the nominal range of applicability of the relations, and secondly,  
recent work have suggested that K-dwarfs experience a stall in their spindown \citep{2020arXiv201002272C}, 
so that such stars can be considerably older than their observed rotation or activity may otherwise suggest.

\subsection {Population membership}
\label{sec:popmem}
The kinematics of this high proper motion star can be used to compute probabilities of membership in different 
populations in the Galaxy. Using the data in 
Table~\ref{tab:stellar} and the methodology of \citet{Reddy2006}, we find galactic velocity components of 
$U = -43.42$ \kmps, $V = -45.96$ \kmps, $W = -30.95$ \kmps. We converted these velocities to the local standard of rest of the Sun to 
$U_{lsr} = -33.42 \pm 0.16$ \kmps, $V_{lsr} = -40.66 \pm 0.12$ \kmps and $W_{lsr} = -23.75 \pm 0.15$ \kmps. This results in a 
probability of the star belonging to the thin disk population of $P(thin) = 0.95 \pm 0.02$, and to the thick disk  
$P(thick) = 0.0516 \pm 0.0002$, and   a vanishingly low probability of the star being old 
enough to belong to the halo population.
The thin disk of the Galaxy is expected to have formed $8.8 \pm 1.7$ Gyr ago \citep{delPeloso2005} which is consistent with the derived ages. 

\begin{table}
 \caption{Spectroscopic  parameters derived with  {\tt {SME}} and {\tt {SpecMatch-Emp}} compared to the stellar effective temperature from Gaia.   
\label{tab:comparison stellar spectroscopic parameters}}
\begin{tabular}{lcccc}
 \hline
     \noalign{\smallskip}
Method  & \teff   & \feh   &   \logg & \vsini   \\  
& (K)  & &(cgs) &(\kms)  \\
    \noalign{\smallskip}
     \hline
\noalign{\smallskip} 
 {\tt {SME}}$^a$  & \steffsme       &          \sfehsme&   \sloggsme & \svsinisme      \\
 {\tt {SpecMatch-Emp}}   &\steffspecm&\sfehpecm&  \ldots    &  \ldots \\
Gaia & $4351^{+204}_{-110}$& \ldots & \ldots & \ldots\\
\noalign{\smallskip} \noalign{\smallskip}
\hline
\noalign{\smallskip}
\multicolumn{5}{l}{$\footnotesize^a$Adopted stellar parameters.}\\
 \end{tabular}
\end{table}

\begin{table}
\centering
 \caption{Stellar mass and radius and the corresponding
 stellar densities derived with different methods and typical mass and radius for an \stype~star.  
\label{tab:comparison stellar parameters}}
\begin{tabular}{lccc}
 \hline
     \noalign{\smallskip}
Method    & \mstar &        \rstar   & $\rho_\star$       \\
  & (\msun)  & (\rsun)  &(\gc)    \\
    \noalign{\smallskip}
     \hline
\noalign{\smallskip} 
{\tt {isochrones}}$^{a, b}$   &\smassisochrones   &\sradiusisochrones &\srhoisochrones   \\
{\tt {Param\,1.5}}$^b$  &  \smassparam  &\sradiusparam  & \sdensityparam     \\
{\tt {SED fitting}}$^b$        &   \smassSEDlogg$^c$ & \sradiusSED &   \ldots   \\
{\tt {SpecMatch-Emp}}    &   \ldots &\sradiusspecm& \ldots      \\
Torres$^{b, d}$ & \smasstorres & \sradiustorres&\sdensitytorres  \\
Boyajian$^e$  &\smassboyajian& \ldots&\ldots  \\
Light curve model$^f$  & \ldots& \ldots& \denstrc[] \\
Spectral type \stype$^g$ & \massstype & \radiusstype & $3.39$\\
\noalign{\smallskip} \noalign{\smallskip}
\hline
\noalign{\smallskip}
\multicolumn{4}{l}{$\footnotesize^a$Adopted stellar mass and radius in the modelling in Sect.~\ref{sec:joint}.}  \\
\multicolumn{4}{l}{$\footnotesize^b$Using \teff, \logg, and \feh~from {\tt {SME}}.}  \\
\multicolumn{4}{l}{$\footnotesize^c$Combining the SED radius with \logg.}  \\
\multicolumn{4}{l}{$\footnotesize^d$\citet{2010AJ....140.1158T} calibration equations.}  \\
\multicolumn{4}{l}{\footnotesize$^e$\citet{2012ApJ...757..112B}  calibration equation from eclipsing binaries using} \\
\multicolumn{4}{l}{\ \ \rstar~from {\tt {isochrones}}.} \\
\multicolumn{4}{l}{\footnotesize$^f$Stellar density obtained from the light curve model (Sect.~\ref{sec:joint}). } \\
\multicolumn{4}{l}{\footnotesize$^g$Typical mass and radius for a   \stype~star.} \\
 \end{tabular}
\end{table}

\section{Joint RV and transit analysis}
\label{sec:joint}

\begin{figure*}
\begin{subfigure}{1\textwidth}
  \centering
  \includegraphics[width=1\linewidth]{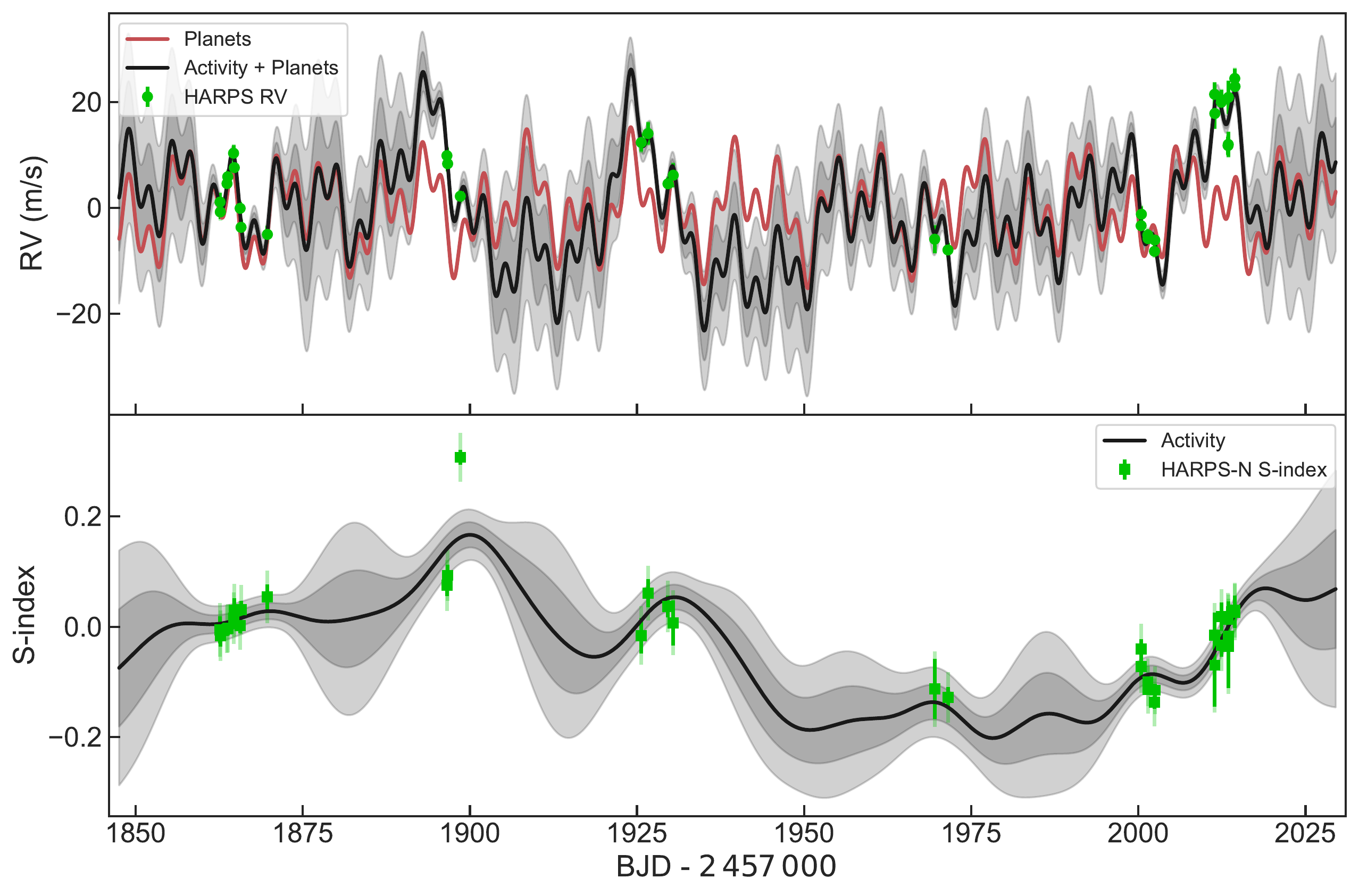} 
  \label{fig:sub-first}
\end{subfigure}
\vfill
\begin{subfigure}{.49\textwidth}
  \centering
  \includegraphics[width=\textwidth]{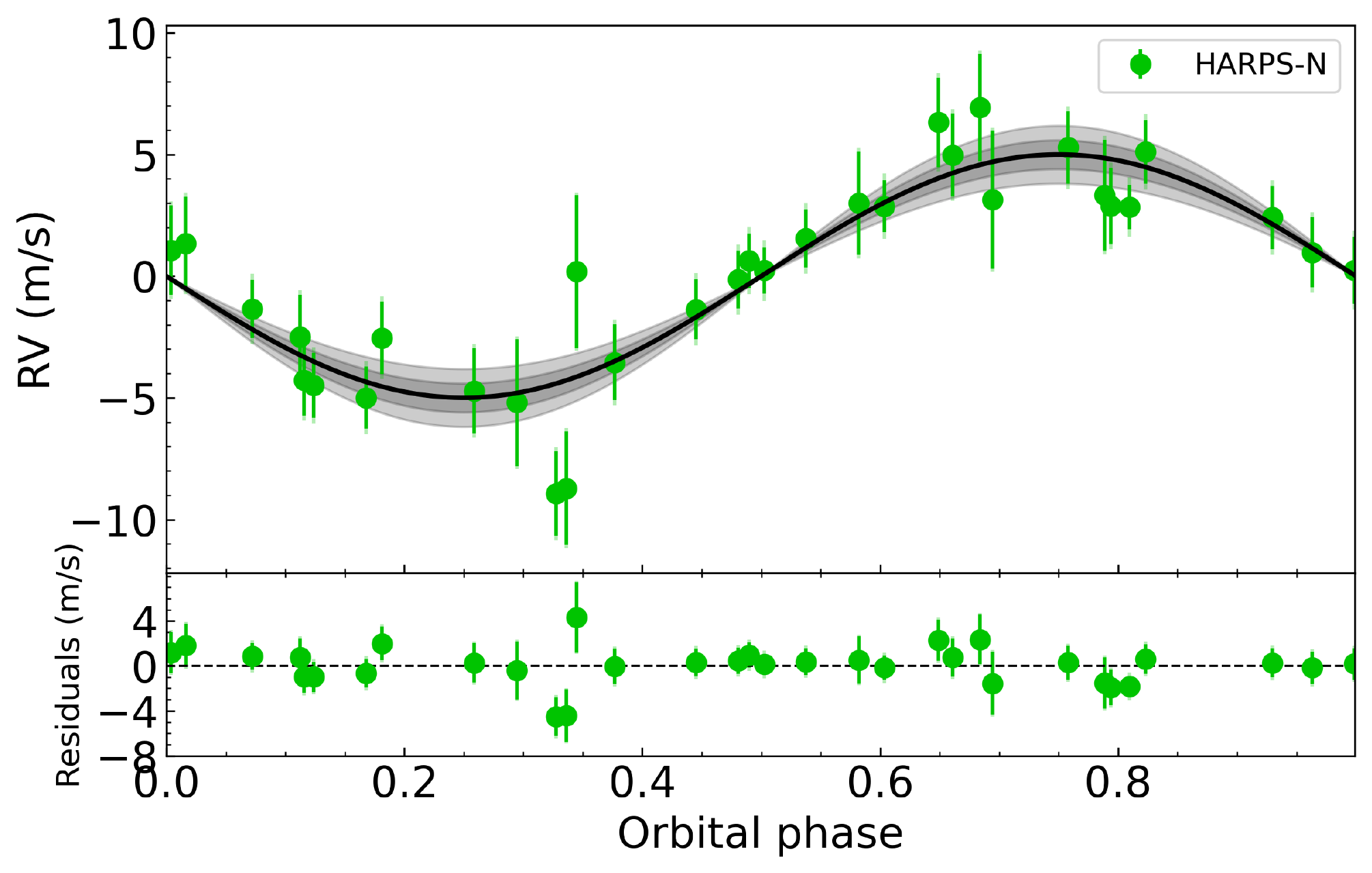}  
  \label{fig:sub-second}
\end{subfigure}
\begin{subfigure}{.49\textwidth}
  \centering
  \includegraphics[width=\textwidth]{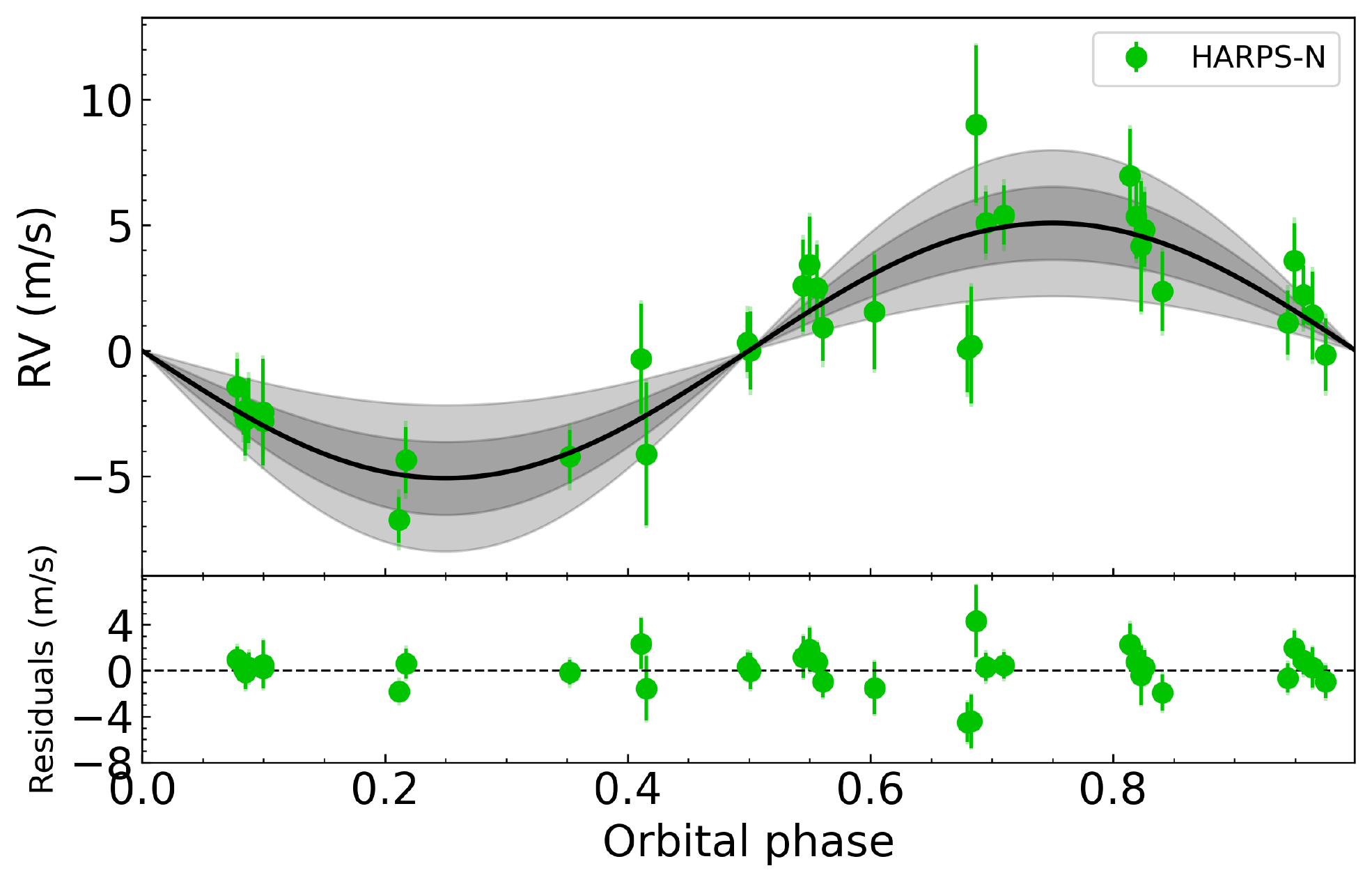}  
  \label{fig:sub-third}
\end{subfigure}
\vfill

\caption{RV (top panel) and S-index (middle panel) time-series. The green markers in each panel represent the HARPS-N RV and S-index measurements with inferred offsets extracted. The solid dark line shows the inferred Multi-GP model, with dark and light shaded areas showing the one and two sigma credible intervals of the corresponding GP model. These regions represent ranges in which other GP curves could also explain the data, with different probability. For the RV panel we also included the RV model for the two planets (solid red line). Bottom panel: HARPS-N RV data folded on the orbital period of each candidate following the subtraction of the systemic velocities, GP signal, and the other planet. The plots also show the inferred RV model for each planet (solid black line) with 1- and 2-sigma credible intervals (shaded areas).
In all the plots the nominal error bars are in green, and the error bars taking into account the jitter ($\sigma_\mathrm{HARPS-N}$) are semi-transparent green. The latter are $ < 1$ \mps\ for the RV data and are hardly visible.}
\label{fig:RVGP}
\end{figure*}

We use the open source software \href{https://github.com/oscaribv/pyaneti2-beta-} \pyan{} \citep{pyaneti}, which uses a Bayesian approach with MCMC sampling for planetary systems parameter estimation, to perform our joint transit and RV analysis, as well as the monotransit and multi-band fits.

Adopting the flattened TESS light curves derived from \citla{} (Sect. \ref{sec:photometry}), together with the LCO single transit data available for 1260.01 (Sect. \ref{sec:photometry}), we model the transits using the \citet{Mandel2002} approach as implemented in \pyan. We sample for the limb darkening parameters utilising the parametrisation $q_1$ and $q_2$ described by \citet[][]{Kipping2013}. Instead of sampling for the scaled semi-major axis, $a/R_\star$, for each candidate, we sampled for the stellar density \rhostar, as parametrized in \pyan.

Section~\ref{Sect: actandrot} describes that our RV measurements contain stellar-induced RV variations. For this reason we use the multidimensional Gaussian-process approach described in \citet{Rajpaul2015} to model our RVs. 
This approach has been used successfully to separate planet signals from stellar activity by e.g. \citet{Barragan2019} and \citet{Mayo2019}.
Briefly, it models RVs together with the activity indicators assuming the same underlying GP, $G(t)$, can describe them. This approach constrains the GP flexibility that could remove planet-induced signals. $G(t)$ can be interpreted as representing the fraction of the visible stellar disc that is covered by active regions at a given time.

For our final GP analysis we model our RVs alongside the S-index as

\begin{equation}
    \begin{matrix}
    \Delta RV & = & V_c G(t) + V_r \dot{G}(t), \\
    \Delta S_{\rm HK} & = & S_c G(t), \\
\end{matrix}
\end{equation}

respectively. 
The variables $V_c$, $V_r$, and $S_c$, are free parameters which relate the individual time series to the Gaussian Process $G(t)$. The RVs depend on the fraction of the stellar disc covered by active regions as well as how these regions move on the surface. For this reason RVs are modelled as a function of $G(t)$ and its time derivative. We use the S-index given that it is an activity indicator that depends on the fraction of the stellar disc covered by active regions, i.e., it can be described by $G(t)$ only. We use the quasi-periodic covariance function 

\begin{equation}
    \gamma(t_i,t_j) = \exp 
    \left[
    - \frac{\sin^2[\pi(t_i - t_j)/P_{\rm GP}]}{2 \lambda_{\rm P}^2}
    - \frac{(t_i - t_j)^2}{2\lambda_{\rm e}^2}
    \right],
    \label{eq:gamma}
\end{equation}

where $P_{\rm GP}$ is the period of the activity signal, $\lambda_p$ the inverse of the harmonic complexity, and $\lambda_e$ is the long term evolution timescale. 

Before committing to a final model setup, we tested different orbital scenarios including two circular orbits, two eccentric orbits, as well as a combination of the two -- inner body with eccentric, outer body with circular orbit, and vice versa. We found that all fits including eccentric orbits provide a solution for the eccentricities consistent with zero. We also calculated the commonly used Bayesian Information Criterion (BIC) and found that the case of two circular orbits is strongly favoured with a $\Delta {\rm BIC} = 15$ better than the second best model. 
This is also consistent with short circularization timescales for short-period planets as well as the \citet{VanEylen19} finding that multi-planet systems tend to feature low eccentricities. Since both candidates are in close-in orbits, the circular case for both yields a value for the stellar density most consistent with the spectroscopically derived one, and given that the current data does not favour the solution with eccentric orbits, we use the circular orbits case scenario as our final model.

Using the above setup and the RVs from {\tt {serval}}, we ran our final model with 500 chains to sample the parameter space. For the burn-in phase we used the last 5000 of the converged chains with a thin factor of 10, leading to a final number of 250,000 independent points for each sampled parameter.

As an additional test we ran a joint model without accounting for the stellar signal in any way. We find that the two planets are still detected, but the HARPS-N jitter is significantly higher (8.8~\mps) than the nominal night-to-night variation ($\approx$0.8~\mps). This points to the presence of additional signals not accounted for by this model. Nevertheless, the results of this test agree within $1\sigma$, thus lending confidence in our choice of final model.

To ensure that our detection is not due to an artefact of the RV data reduction, as an extra check we performed our final model setup using the DRS-derived RVs. The results once again agree to within $1\sigma$ of our adopted parameters.

Lastly, to check that our results do not depend on the sampling algorithm, we used the code \texttt{juliet} \citep{juliet} to model jointly the photometric and Doppler data. The algorithm is built on many publicly available tools for the modeling of transits \citep[\texttt{batman},][]{batman}, RVs \citep[\texttt{radvel},][]{radvel}, and GP (\texttt{george}, \citealt{Ambi2016}; \texttt{celerite}, \citealt{celerite}), and computes efficiently the Bayesian log-evidence using the importance nested sampling included in the \texttt{dynesty} package \citep{dynesty}. We use the same set of priors presented in Table~\ref{params}, but for the GP we use an exponential-sine-squared kernel of the form $k_{i,j} = \sigma^2_\mathrm{GP,RV} \exp\left(- \alpha_\mathrm{GP,RV} (t_i - t_j)^2 - \Gamma_\mathrm{GP,RV} \sin^2 \left[\frac{\pi |t_i - t_j|}{P_{\rm rot;GP,RV}}\right]\right)$ with a uniform prior in $P_{\rm rot;GP,RV}$ ranging from 22 to 43\,d. The \texttt{juliet} package does not have the possibility to perform fits with multi-dimensional GP so in this case we apply it only on the RV data. Nevertheless, the fitted parameters from the joint fit with \texttt{juliet} are in perfect agreement with the results from \texttt{pyaneti}, confirming the robustness of the different analyses and the derived orbital parameters.

A summary of our results, including the fitted parameters and priors are presented in Table~\ref{params}. Figure~\ref{fig:RVGP} shows the RV and S-index timeseries together with the inferred models.
It should be noted that in Fig.~\ref{fig:RVGP} the uncertainties of the inferred models (shadow regions) are relatively large, which is caused by the sub-optimal sampling of the data and the flexibility of the GP model.
This figure illustrates the usefulness of the multi-dimensional GP used in this work as it is clear how the RV GP model is constrained by the changes in the S-index \citep[similar to Fig.~2 of][]{Barragan2019}.

\begin{figure}
\centering
	\includegraphics[width=1\linewidth]{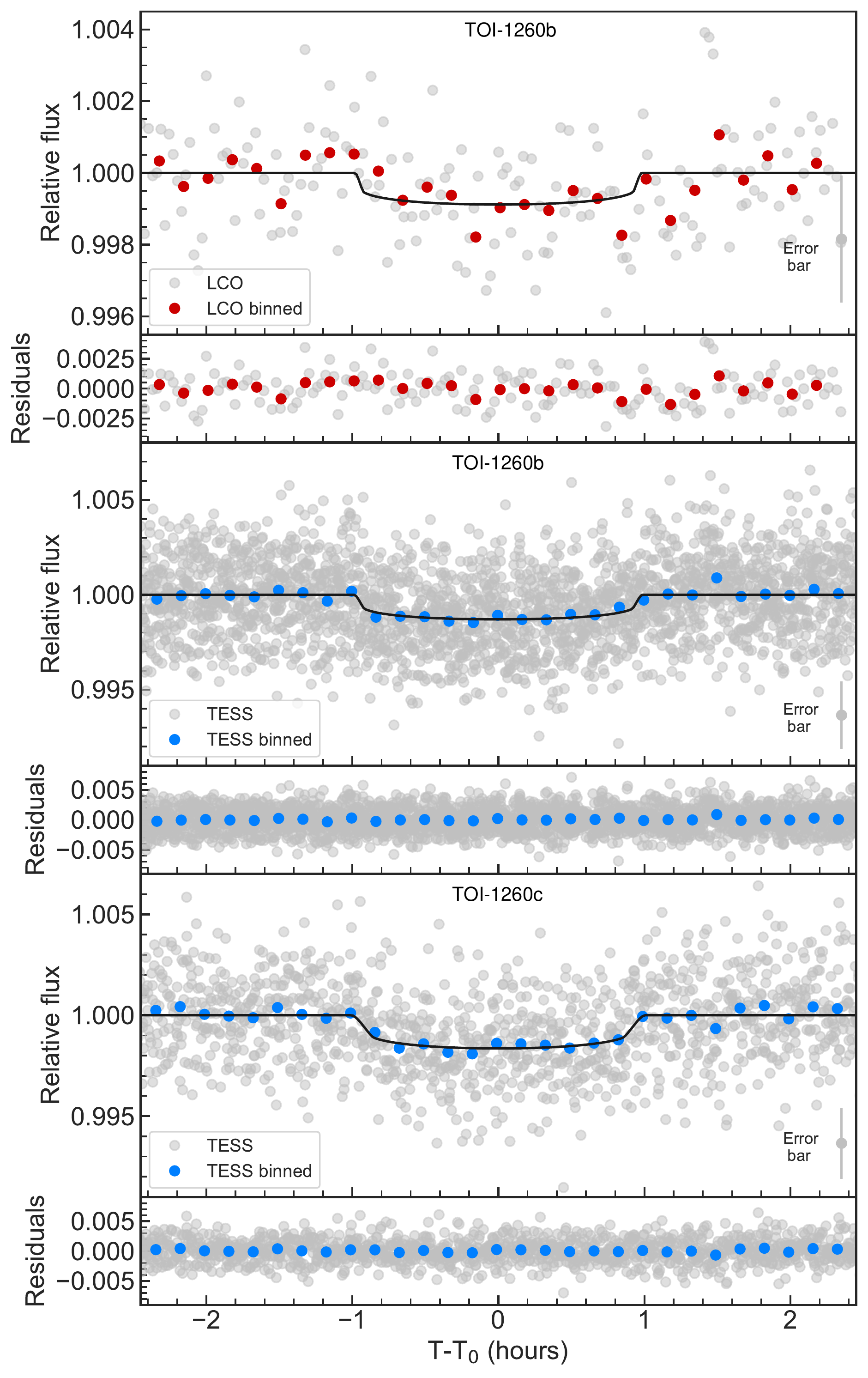}
    \caption{\planetb{} and \planetc{} transits. The panels show a flattened LCOGT and TESS light curves with residuals folded to the orbital periods of the planets. Black lines show the best-fitting transit models. The LCO and TESS radius estimates for planet b agree to nearly 1$\sigma$. Data are shown in the nominal 2-min cadence mode and binned to 10 min. Typical error bar for nominal data is shown at the bottom right for each panel.}
    \label{fig:bands}
\end{figure}

Figure~\ref{fig:RVGP} also shows phase-folded RV data of planets b (1260.01) and c (1260.02) together with the corresponding inferred RV model, while Fig.~\ref{fig:bands} shows the single transit event of planet b detected by LCO as well as the phasefolded transits of both planets as obtained from TESS photometry.

\begin{table*}
  \caption{Summary of the system parameters from the stellar modelling in Sect.~\ref{sec:stellar} and the joint RV and transit modelling with \pyan\ in Sect.~\ref{sec:joint}. \label{params}}  
  \begin{tabular}{lcc}
  \hline
  Parameter & Prior$^{(\mathrm{a})}$ & Value$^{(\mathrm{b})}$  \\
  \hline
  \multicolumn{3}{l}{\emph{ \bf Model Parameters for  \target}} \\
  \hline
  \multicolumn{3}{l}{\emph{\planetb}} \\
    \noalign{\smallskip}
    Orbital period $P_{\mathrm{orb}}$ (days)  &  $\mathcal{U}[3.1270 , 3.1280]$ & \Pb[] \\
    Transit epoch $T_0$ (BJD - 2,450,000)  & $\mathcal{U}[  8684.0050 , 8684.0250 ]$ & \Tzerob[]  \\
    $e$  &  $\mathcal{F}[0]$ & 0  \\
    $\omega_\star $  &  $\mathcal{F}[\pi/2]$ & $\pi/2$  \\
    Scaled planetary radius $R_\mathrm{p}/R_{\star}$ &  $\mathcal{U}[0.01 , 0.10]$ & \rrb[]  \\
    Impact parameter, $b$ &  $\mathcal{U}[0,1]$  & \bb[] \\
    Radial velocity semi-amplitude variation $K$ (m s$^{-1}$) &  $\mathcal{U}[0, 25]$ & \kb[] \\
    \hline
    \multicolumn{3}{l}{\emph{\planetc}} \\
    \noalign{\smallskip}
    Orbital period $P_{\mathrm{orb}}$ (days)  &  $\mathcal{U}[7.4925 , 7.4940]$ & \Pc[] \\
    Transit epoch $T_0$ (BJD - 2,450,000)  & $\mathcal{U}[  8686.1050 , 8686.1300 ]$ & \Tzeroc[]  \\
    $e$  &  $\mathcal{F}[0]$ & 0  \\
    $\omega_\star $  &  $\mathcal{F}[\pi/2]$ & $\pi/2$  \\
    Scaled planetary radius $R_\mathrm{p}/R_{\star}$ &  $\mathcal{U}[0.01 , 0.10]$ & \rrc[]  \\
    Impact parameter, $b$ &  $\mathcal{U}[0,1]$  & \bc[] \\
    Radial velocity semi-amplitude variation $K$ (m s$^{-1}$) &  $\mathcal{U}[0, 25]$ & \kc[] \\
    \hline
    GP Period $P_{\rm GP}$ (days) &  $\mathcal{U}[22,43]$ & \jPGP[] \\
    $\lambda_{\rm P}$ &  $\mathcal{U}[0.1,5]$ &  \jlp[] \\
    $\lambda_{\rm e}$ (days) &  $\mathcal{U}[1,200]$ &  \jle[] \\
    $V_{c}$ (\kms)  &  $\mathcal{U}[0,0.1]$ & \jArvc \\
    $V_{r}$ (\kms) &  $\mathcal{U}[0,1]$ & \jArvr \\
    $S_{c}$  &  $\mathcal{U}[0,1]$ & \jAsmwc \\
    Offset HARPS-N (\kms) & $\mathcal{U}[ -0.05 , 0.05 ]$ & \HARPSN[] \\
    Offset \smw \ & $\mathcal{U}[0.5 , 1.9 ]$ & \Smw[]  \\
    Jitter term $\sigma_{\rm HARPS-N}$ (\ms) & $\mathcal{J}[10^{-3} , 10^{-1}]$ & \jHARPSN[] \\
    Jitter term $\sigma_{\rm S-index}$ & $\mathcal{J}[10^{-3} , 10^{-1}]$ & \jSmw[] \\
    Limb darkening $q_1$, \tess\ & $\mathcal{U}[0,1]$ & \qone \\
    Limb darkening $q_2$, \tess\ & $\mathcal{U}[0,1]$ & \qtwo \\
    Limb darkening $q_1$, LCO & $\mathcal{U}[0,1]$ & \qoneLCO \\
    Limb darkening $q_2$, LCO & $\mathcal{U}[0,1]$ & \qtwoLCO \\
    Jitter term $\sigma_{\tess}$ ($\times 10^{-6}$) & $\mathcal{U}[0,1 \times10^{3}]$ & \jtr \\
    Jitter term $\sigma_{\rm LCO}$ ($\times 10^{-6}$) & $\mathcal{U}[0,1 \times10^{3}]$ & \jtrLCO \\
    Stellar density $\rho_{\star}$ (${\rm g\,cm^{-3}}$) &   $\mathcal{U} [ 0.1, 10 ]$ & \rhostellar[] \\
    \hline
    \hline
    \noalign{\smallskip}
    {\emph{\bf Derived parameters}} & {\emph{\bf \planetb}} & {\emph{\bf \planetc}} \\
  \noalign{\smallskip}
    Planet mass ($M_{\oplus}$)  & \mpb[] & \mpc[] \\
    Planet radius ($R_{\oplus}$)  & \rpb[] & \rpc[] \\
    Planet density (${\rm g\,cm^{-3}}$) & \denpb[] & \denpc[] \\
    Scaled semi-major axis $a/R_\star$ & \arb[] & \arc[] \\
    Semi-major axis $a$ (AU)  & \ab[] & \ac[] \\
    Orbital inclination $i$ (deg)  & \ib[] & \ic[] \\
    Transit duration $t_{\rm tot}$ (hours) & \ttotb[] & \ttotc[] \\
    Equilibrium temperature $^{(\mathrm{c})}$ $T_{\rm eq}$ (K) & \Teqb[] & \Teqc[] \\
    Insolation $F_{\rm p}$ ($F_{\oplus}$)   & \insolationb[] & \insolationc[] \\
    Planet surface gravity$^{(\mathrm{d})}$ (cm\,s$^{-2}$) & \grapb[] & \grapc[] \\
    Planet surface gravity (cm\,s$^{-2}$)  & \grapparsb[] & \grapparsc[] \\
    \hline
    \hline
    \multicolumn{3}{l}{\emph{\bf Adopted stellar parameters}} \\
    \noalign{\smallskip}
    Stellar mass (\msun)  & $\cdots$ & \smassisochrones \\
    Stellar radius (\rsun)  & $\cdots$ & \sradiusisochrones \\
    Stellar density (\gc) & $\cdots$ &\srhoisochrones \\
    Effective temperature ($K$) & $\cdots$ & \steffsme[] \\
    Bolometric luminosity (\lsun) & $\cdots$ & \Lisochrones \\
    \hline
   \noalign{\smallskip}
  \end{tabular}
  \begin{tablenotes}\footnotesize
  \item \emph{Note} -- $^{(\mathrm{a})}$ $\mathcal{U}[a,b]$ refers to uniform priors between $a$ and $b$, $\mathcal{J}[a,b]$ to modified Jeffrey's priors calculated using eq. 16 in \citet[][]{Gregory2005}, and $\mathcal{F}[a]$ to a fixed value $a$.  
  $^{(\mathrm{b})}$  Inferred parameters and errors are defined as the median and 68.3\% credible interval of the posterior distribution.
 {  $^{(\mathrm{c})}$ Assuming an albedo of 0.
 $^{(\mathrm{d})}$ Calculated from the scaled-parameters as in \citet[][]{Southworth2007}.}
\end{tablenotes}
\end{table*}

\begin{figure}
\centering
	\includegraphics[width=1\linewidth]{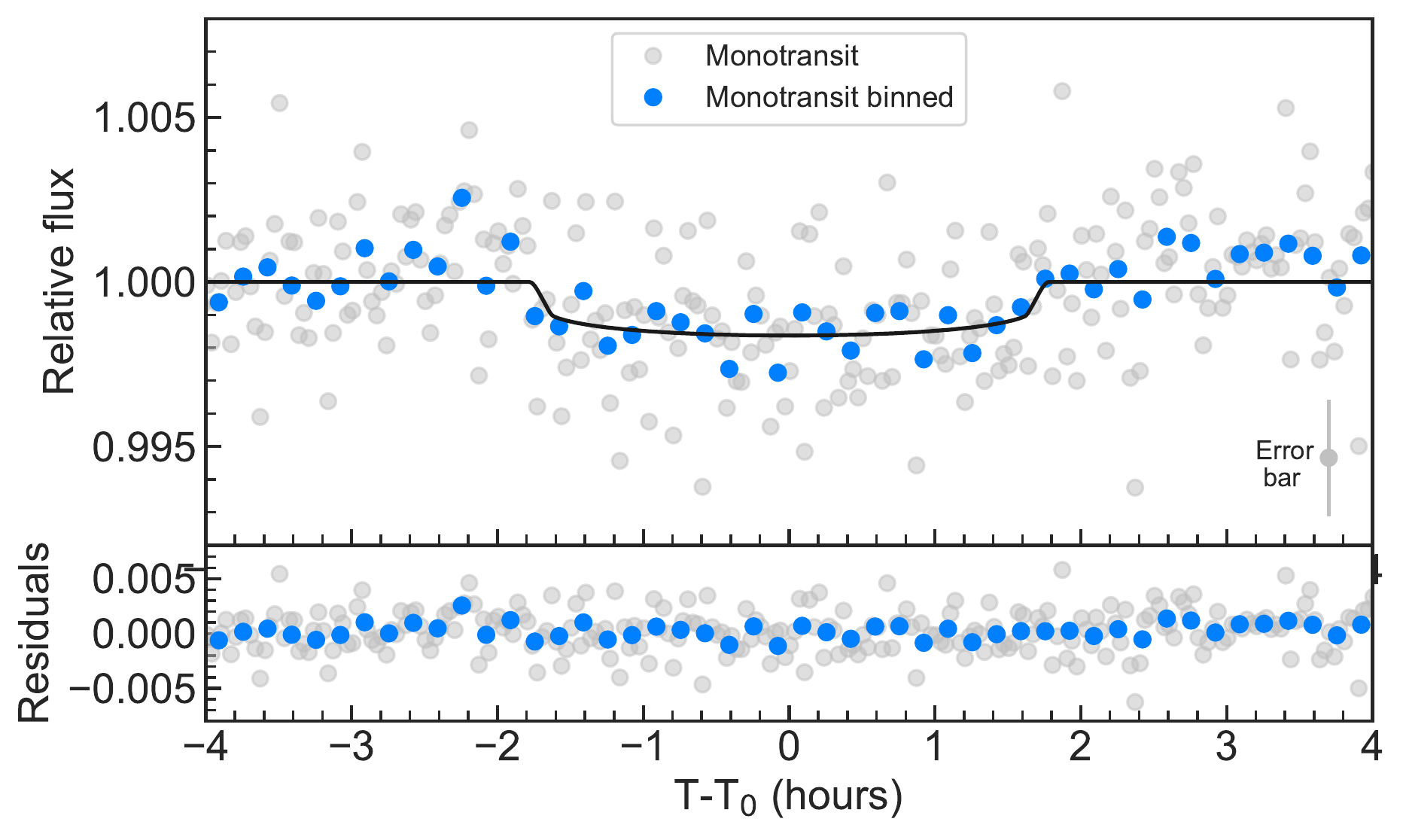}
    \caption{The single transit of the tentative outer planet d seen in Sector 21. The \pyan\ transit model yields a $T_0$ of \Tzerod\ and a depth of \depthd, which corresponds to a radius of \rpd. Data are shown in the nominal 2-min cadence mode and binned to 10 min, with typical error bar for nominal data in the bottom right.}
    \label{fig:mono}
\end{figure}

\subsection{Tentative outer planet}

\label{sec:Planetd}

As discussed in Sect.~\ref{sec:photometry}, we report an additional transit-like event in Sector 21. A counterpart of this feature is not visible in Sector 14, although it is possible that the transit occurred during the \char`\~1-day data gap between orbits (Fig.~\ref{fig:lc_all}). This transit-like feature does not coincide with a spacecraft momentum dump. 

To model the monotransit, we again turn to \pyan. We follow a similar approach as in \citet{Osborn2016}. Assuming a circular orbit and based on the transit shape, our single-transit model (Fig.~\ref{fig:mono}) gives a range of physically possible periods of [13.4, 56.3], a transit depth of \depthd, which in turn yields a radius of \rpd. We further narrowed down the period range based on the length of TESS observations and the apparent lack of occurrence of another such transit event during the observing windows. Our final possible periods are listed in Table~\ref{mono_P}.
The binned and unbinned transit data and inferred model of the aforementioned monotransit visible in Sector 21 are displayed in Fig.~\ref{fig:mono}.

In an attempt to try and explore further the physical properties of this tentative outer planet, we performed an MCMC analysis identical to our adopted one, but we added an extra planetary signal with ephemeris corresponding to the transit of the tentative planet d. We used a prior on T$_0$ of [8879.2, 8879.4], and a wide prior on the period of [20.0, 70.0] and created marginalized posterior distributions using \pyan. We were unable to further constrain the period but we found the maximum allowed semi-amplitude to be 18.4~\mps (99\% confidence interval).

We cannot constrain this further as there is also no sign of another planet in our RV dataset. However, with a maximum semi-amplitude of 18.4~\mps, this translates to a mass of 76.3~\mearth. Therefore, if the signal at 1879.32 days is caused by a transiting object, this object belongs to the planetary mass domain.

\begin{figure}
\centering
	\includegraphics[width=1\linewidth]{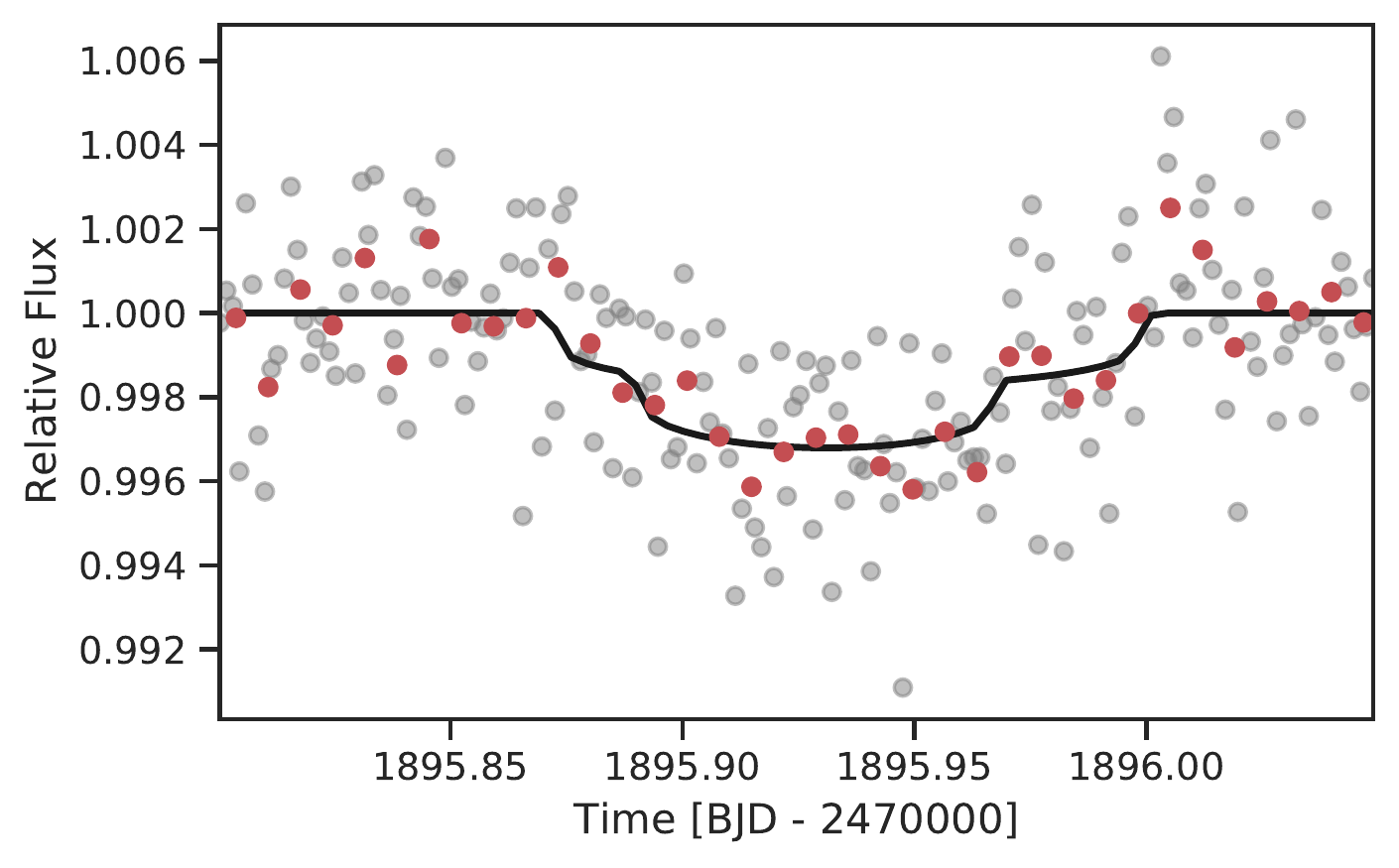}
    \caption{Two-transit model of the potentially overlapping transits of planet c and the tentative planet d, around 1895.95 BJD - 2457000. Gray circles show the flattened TESS data, with ten-minute bins as red circles, and solid line showing the inferred transit model including both planet signals.}
    \label{fig:due}
\end{figure}

We further note that the minimum period shown in Table~\ref{mono_P} is 20.3 days. This constraint comes from the minimum period that the tentative outer planet has to have in order to not be observed transiting again in the light curve. We however, note that there is a transit of \planetc\ between the range 8895.80-8896.05 BJD - 2450000 that looks significantly deeper. This can be caused by some unknown systematics in the light curve or another obscuring object. To investigate this, we performed a simple model adding an extra single transit to a model of planet c between the range  8895.80-8896.05 BJD - 2450000. We thus found that we obtain a better model to the data if we add a signal with a time of mid-transit of $8895.938 \pm 0.005$, depth of $1705 \pm 350$ ppm, and transit duration of $2.9 \pm 0.3$ hours. Figure \ref{fig:due} shows a plot with the two-transit model. These tentative transit parameters are consistent within 2-sigma with our single transit event at 8879.3210683 BJD - 2450000. If this detected signal is real and it corresponds to a second transit of the tentative planet d, then its period would be $\sim 16.61$ days (see Sect.~\ref{sec:photometry}). Unfortunately, with this period, the only other visible transit in the available TESS light curves would have fallen in the data gap of Sector 14.

We then repeat a similar approach as the one described in Sect.~\ref{sec:joint}, with an extra Keplerian signal with a tight prior on the ephemeris of the tentative $16.61$-day planet but we have no clear detection of a RV signal at that period. If this planet is real, based on this three-planet model, its period, radius and transit duration would be $16.613 _{ - 0.006 } ^ { + 0.008 }$ days and $2.75 _{ - 0.177 } ^ { + 0.172 }$~\rearth\ and $3.11 _{ - 0.15 } ^ { + 0.20 }$ hours, respectively. 
The 99\% credible interval for the maximum semi-amplitude would be around $13$~\mps, which in turn translates to a maximum mass of around $39$~\mearth. Adding this signal has an insignificant effect on the parameters of planet b, while planet c shows a slight decrease in radius to $2.68 \pm 0.14$~\rearth\ and an increase in mass to $13.39 _{ - 3.26 } ^ { + 3.49 }$~\mearth. Both of these agree well with our officially reported estimates in Table~\ref{params}.

Based on these results, we take a conservative approach and we conclude that, based on the available information, we cannot claim a planet with a period of $16.61$ days. However, if there is such a planet, it could be confirmed by photometric ground or space-based follow-up. Fortunately, TESS will observe \target\ again in three more sectors  -- 41, 47 and 48. We note, however, that a RV follow-up would be more challenging because this tentative period is close to half the rotation period of the star.

\section{Discussion}
\label{sec:discussion}

\subsection{Dynamical stability}
\label{DynStab}

The dynamical viability of multi-planet systems is an important component of assessing valid architectures. Testing dynamical integrity and subsequent orbital evolution has played a key role in understanding Kepler systems \citep{lissauer2011,li2014,kane2015b,kane2019c}. To test the stability of the orbital solution for our two confirmed planets in the \target\ system, we executed N-body integrations using the Mercury Integrator Package \citep{chambers1999}. We adopted the stellar, planetary masses and semi-major axes from Table~\ref{params}. We further assumed initial circular orbits for both of the planets. The simulation was performed for $10^7$ simulation years with a time step of 0.1 days to properly sample the relatively short orbital period of the inner planet. The results of the simulation showed no signs of instability, and the eccentricities of both planets remained below $10^{-3}$ for the duration of the simulation. This demonstrates that the gravitational well of the star is the overwhelmingly dominant influence on the planetary dynamics within their compact system configuration.
Given the proximity of the planets to each other, we also investigated the possibility of determining upper mass limits that retain dynamical stability. We gradually increased the masses of both planets independently until the dynamical integrity of the system was compromised during a series of $10^6$ year simulations. These simulations showed that the maximum masses for both planets are loosely constrained based on their dynamical interactions, with maximum masses approaching several Jupiter masses before significant instability occurs.

Tipped off by the suspected presence of an outer planet, we decided to check if the system exhibits Transit Timing Variations (TTVs). We performed a TTV analysis using \texttt{PyTTV} \citep[Python Tool for Transit Variations,][]{kups11289}, which showed that a linear ephemeris can be fit between the sectors. The ephemerides from our modelling results (Table~\ref{params}) and the lack of TTVs allows for future observations of the system using other facilities to be scheduled efficiently.

\subsection{Characterization of the \target\ planets}

\begin{figure}
\centering
	\includegraphics[width=1\linewidth]{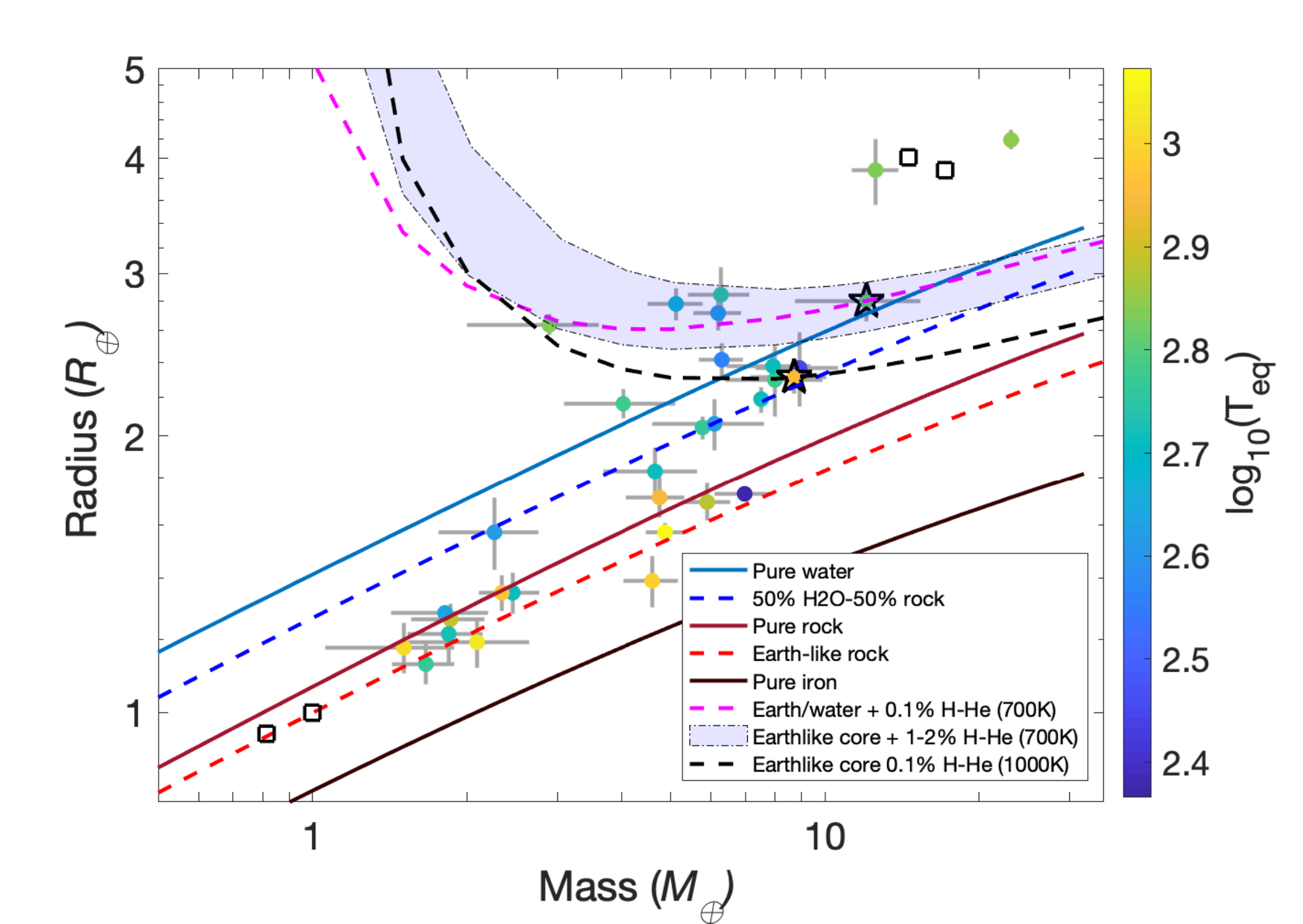}
    \caption{Mass-radius diagram of planets with measured masses better than 30\% and radii better than 10\% orbiting mid-M to mid-K dwarfs (3000 -- 4400 K). In total there are 26 planets in 19 multi-planet systems. Models of core compositions without atmosphere \citep{Zeng2016} and with atmosphere \citep{Zeng2019} at different equilibrium temperatures are also plotted. \planetb\ and \planetc\ are marked with star symbols, and squares are the Solar system planets. }
    \label{fig:MR_MK}
\end{figure}

\begin{figure}
\centering
	\includegraphics[width=1\linewidth]{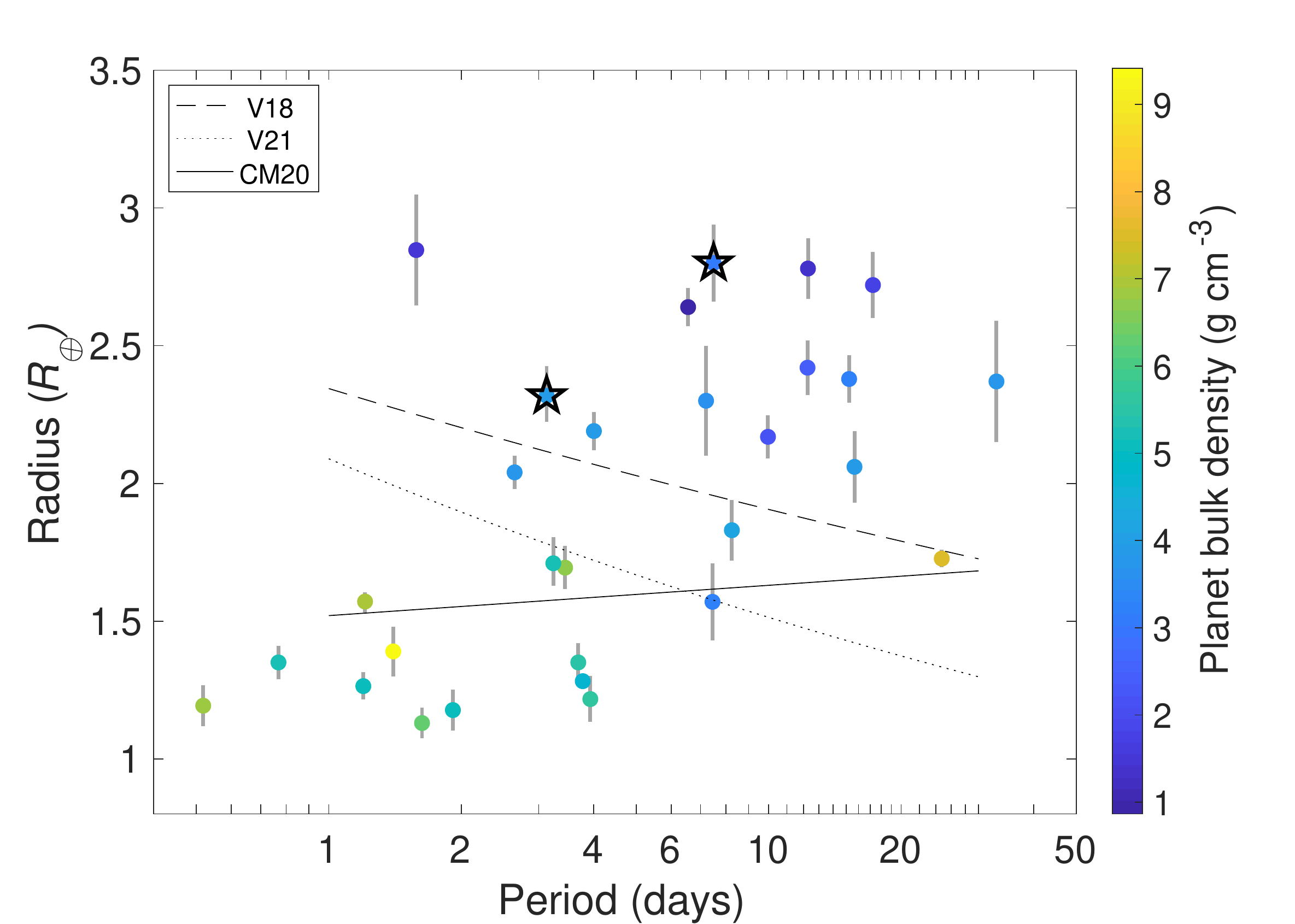}
    \caption{Radius-period diagram for the same planet population as in Fig.~\ref{fig:MR_MK}. The dashed line is the fit to the FGK radius valley from V18 \citep{VanEylen2018}, the solid line refers to stars $\leq$ 4700K as per CM20, \citet{Cloutier2020}, while the dotted line -- to M dwarfs with $T_\mathrm{eff} < 4000$~K \citep{VanEylen21}. \planetb\ and \planetc\ are again marked with star symbols. Planet c is found comfortably above all three radius valleys fits, while planet b lies on the edge of the V18 fit. }
    \label{fig:RP_diagram}
\end{figure}

Two important factors that influence the radius distribution of planets are the semi-major axis and the mass of the host star \citep{FulPet18, Wu19, Cloutier2020, VanEylen21}. Both of these determine a planet’s X-ray/UV irradiation evolution. Since the magnitude and evolution of the X-ray luminosity differs between sun-like and low mass stars \citep{2019ApJ...876...22M, 2021A&A...645A..41L}, we show in Fig.~\ref{fig:MR_MK} a mass-radius diagram with planets orbiting mid-M to mid-K stars (here defined as having \teff~between 3000 and 4400~K) 
measured to a precision better than $30\%$ in mass and $10\%$ in radius. 
We also plot theoretical models of planet core compositions without an atmosphere \citep{Zeng2016} and with an atmosphere \citep{Zeng2019} at different equilibrium temperatures matching \planetb{} and \planetc.
From Fig.~\ref{fig:MR_MK} we see that the two mini-Neptunes in the \target~system may be water worlds or rocky worlds with H-He atmospheres inflating their radii. 
The position of \planetb{} in the diagram is consistent with a planet composition of $50\%$ Earth-like rocky core (32.5\% Fe + 67.5\% MgSiO$_3$) and $50\%$ H$_2$O ice without an atmosphere, or an Earth-like rocky core with a H-He atmosphere of $\sim0.1\%$.
The position of \planetc{}, with \mpc, \rpc, and a bulk density of \denpc, lies above the pure water line in the diagram. The orbital period and equilibrium temperature are 7.493~days and 643~K, respectively. We find that two models fit the position in the diagram: an Earth-like rocky core with a H-He atmosphere of $\sim2\%$, or alternatively, a core composed of a mix of $49.95\%$ rock and $49.95\%$ ices and a H-He atmosphere of $\sim0.1\%$. 

Since the location of the photoevaporation valley is a function of stellar mass and is thus different for low-mass vs solar-type stars, we plot in Fig.~\ref{fig:RP_diagram} the same \teff\ ranges as in  Fig.~\ref{fig:MR_MK}. As evident from Fig.~\ref{fig:RP_diagram}, both \target\ planets lie above the photoevaporation gap \citep{VanEylen2018, Cloutier2020, VanEylen21}, or close to its edge as is the case of planet b. Depending on the photoevaporation valley fit used, however, planet b could also lie exactly in the transition zone \citep{Wu19}. It should be noted that the \citet{VanEylen2018} curve is based on hotter (4700 – 6500 K) and thus higher mass stars, the \citet{Cloutier2020} and \citet{Wu19} curves relate to low mass stars (mid-K and cooler), while the \citet{VanEylen21} refers to M dwarfs with $T_\mathrm{eff} < 4000$~K.
We have color-coded the planet bulk densities in Fig.~\ref{fig:RP_diagram}, and it is evident that the planets above the radius gap have lower densities than the planets below. The \target\ planets are consistent with this trend as they have relatively low densities and their compositions are degenerated. They both are consistent with both (a) an Earth-like composition of iron and silicates, and (b) an Earth-like core with a substantial fraction of water ice. We delve into possible reasons for this ambiguity in the following sections.

\subsubsection{Mass and radius evolution induced by photoevaporation}

\begin{figure}
\centering
	\includegraphics[width=1\linewidth]{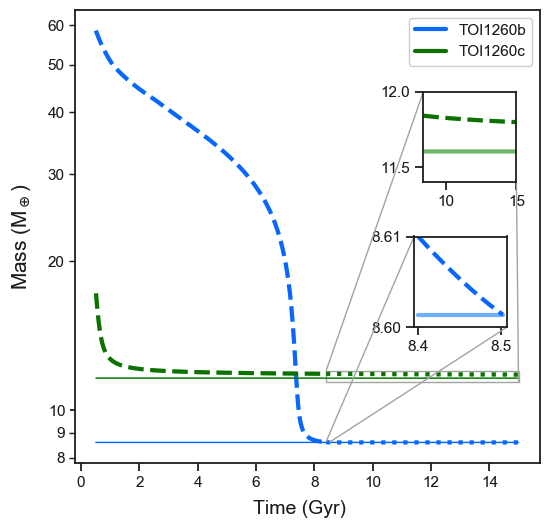}
    \caption{Mass temporal evolution of the \target\ planets assuming a nominal stellar age of 8.4 Gyr, a rock/metal core and a H-He envelope of 0.1\% and 2\% for planets b and c, respectively. Dotted lines refer to the evolution from the current age of the system until 15 Gyr. Dashed lines refer to the inferred evolution from early to current times. The insets show a zoomed-in view of the future evolution, where the semi-transparent solid lines denote the core mass of each planet. It can be clearly seen that planet b would lose a 0.1\% H-He atmosphere in about 100 Myr, while the atmosphere of planet c is stable against photoevaporation.}
    \label{fig:massev}
\end{figure}

\begin{figure}
\centering
	\includegraphics[width=1\linewidth]{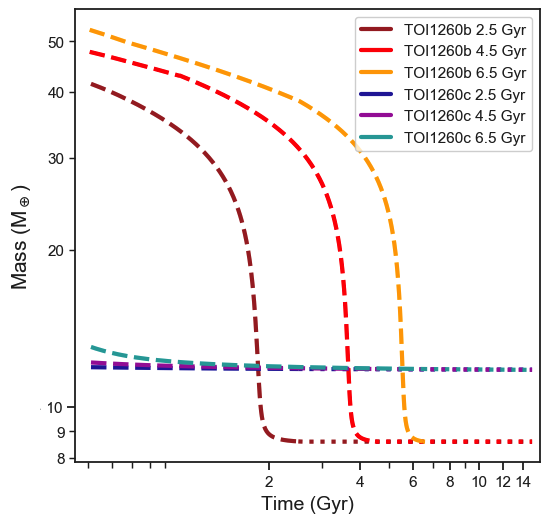}
    \caption{Mass temporal evolution of the \target\ planets as per Fig.~\ref{fig:massev} but considering different stellar ages.}
    \label{fig:ages}
\end{figure}

In order to shed light on which planet composition model \planetb\ and \planetc\ belong to, we investigate the mass and radius temporal evolution induced by atmospheric photoevaporation. To this end, we study the temporal evolution of the high-energy stellar radiation and the planetary radius. We consider a primary H-He atmosphere, a rock/iron core as per \citet{LopFor14}, assume circular orbits, ignore any migration effects and follow the hydrodynamic-based approximation developed by \citet{Kubyshkina2018}. A major driver behind atmospheric hydrodynamic mass loss is the X-ray luminosity since X-ray heating from the star can drive a system to an intense hydrodynamic escape phase \citep{Er07,2008A&A...477..309P,2019A&A...624A.101L}. We estimated the current X-ray luminosity using the ${\rm log}(R'_{\rm HK})$, our SED bolometric luminosity and the relationships in \citet{Houdebine17}, obtaining $L_{\rm X}=4.51\times10^{27}$ erg s$^{-1}$. Since the evolution of extreme ultraviolet radiation follows the evolution of X-ray radiation, we accounted for the X-ray luminosity evolution by using the prescriptions given in \citet{Penz08a} and the relation given in \citet{SF11}.
Following \citet{Pop21}, we account for the evolution of the planetary radius by means of the analytic fit given in \citet{LopFor14}. The analytic fit provides the radius envelope, \renv, as a function, among other parameters, of the atmospheric mass fraction, \fatm, and the age of the system, which in turn allows us to also account for gravitational shrinking.

Calculating the planetary mass (\Mp), \fatm\ and \renv\ is an iterative process. As a first step, we look at the future evolution of the system from its present age ($\sim8.4$~Gyr)
to 15~Gyr and assume \fatm\ values of 0.1\% and 2\% for
planets~b and c, respectively. These correspond to the composition scenarios of an
Earth-like rocky core with a H-He envelope for both planets. We then calculate the
corresponding \renv\ and estimate the core radius simply as the difference between the
measured by photometry planetary radius, \rp\ (Table~\ref{params}) and the calculated
\renv. Next, we updated \fatm\ and \Mp\ at each time step according to the mass loss and used them to calculate a new \renv, adding the latter to the core radius, to finally obtain
the new \rp. We find that planet b loses its atmosphere in about 100~Myr, while planet~c
retains part of it until the end of the run.

\subsubsection{Effect of the stellar age}

To better understand the situation, we take this analysis one step further by tracing the system’s evolution back in time. Assuming the aforementioned scenarios, since the core does not change in size or mass, we create a synthetic population of planets and assign to them the current core radii and masses of our planets. This leaves \fatm\ to dictate the total mass, while the total radius is again based on the analytic fit by \citet{LopFor14}. We then looked at the planets that ended up with a similar current mass, radius and \fatm\ and looked at their predicted past histories.

Figure~\ref{fig:massev} shows the result of both the future (dotted lines) and past (dashed) simulation runs. We trace the planets back to 0.5 Gyr from the assumed birth of the system  and see that in the case of planet b (purple curve), we reach a mass of nearly 60~\mearth\footnote{The hydrodynamic-based approximation works in the 1 -- 39~\mearth\ mass range, so beyond this limit we use the energy limited approximation by \citet{Er07} to model the mass loss.}. 
In the case of planet c (Fig.~\ref{fig:massev}, green curve), we find a much more controlled mass loss process, reaching a starting point of about 17~\mearth. This, and the fact that the future evolution of the atmosphere is stable against evaporation in the long run, makes the Earth-like core with 2\% atmosphere case plausible.

While it is possible to trace the planets further back in time, we stop at 0.5 Gyr since the results beyond that would be subjected to the further uncertainty associated with the stellar rotation rate during the saturation phase early in the star's life.

Due to the uncertainty in the stellar age, we decided to test the same cases as before but with lower age values. We chose ages of 2.5, 4.5 and 6.5 Gyr and reran the models for both planets (Fig.~\ref{fig:ages}). In short, we find that planet b still loses its atmosphere in about the same time frame ($\sim$~100 Myr); planet c retains a long-term stable atmosphere as before and its temporal evolution is almost completely independent of the age of the star. This result is not so surprising when we consider the fact that the X-ray luminosity is most intense in the early evolutionary stages, during which most of the atmospheric mass loss occurs. These results are generally consistent with the above findings for the nominal age, showing that the mass and radius evolution of the planets is robust for a wide range of stellar ages.

We, however, note, that 100 Myr is a short time compared to the overall life of the star, especially if the star is older. This makes it relatively unlikely that we would currently be witnessing the process of planet b losing a primary H-He atmosphere. 

The fact that the nominal age is at the upper limit of the thin disc population age range (see Sect.~\ref{sec:popmem}), as well as the result that the mass evolution of both planets is well consistent with a significantly younger star, suggests the possibility that this star is, in fact, younger, which in turn emphasizes the fact that a high precision of the stellar age estimate can decrease the degeneracy in the determination of planet interiors.

\subsubsection{Planetary composition and atmospheric characterization potential}

Looking back to the two scenarios for \planetb, we consider the composition of a 50\% Earth-like core and 50\% water-ice case, likely mixed rather than layered as suggested by \citet{Vazan2020}, to be more probable. However, the above models do not take into account planet migration or rather assume orbit migration took place quickly (a few Myr) early in the system's history, so a complex migration history could have played a role in this relatively old system. We also note that the X-ray luminosity evolution is calculated using a scaling law just for the mean value \citep{Penz08a} and does not account for different levels of high energy radiation to which planets could be
subjected during their early evolutionary stages. The effects of stellar wind and magnetically-driven cataclysmic events originating from the stellar surface, which could affect the rate of photoevaporation, are also ignored. Furthermore, our simulations only consider the case of H-He primary atmospheres. Thus, our results do not exclude the possibility of secondary envelopes, or primary envelopes of a different composition, which may in turn be smaller and more difficult to lose under atmospheric escape processes. The latter case could mean that \planetb\ and \planetc\ are representatives of a high-metallicity population of hot Neptunes as discussed in depth by \citet{Moses2013}. \citet{Hu15} proposed the existence of He atmosphere planets, and that many sub-Neptune-sized exoplanets in short orbits could possess such atmospheres. They proposed that such an atmosphere could explain for example the emission and transmission spectra of GJ436b. While much smaller and less massive than GJ436b, \planetb\ has a similar orbital period and equilibrium temperature, and could be a firm candidate to posses a He atmosphere. Those atmospheres contain trace amounts of hydrogen, carbon, and oxygen, with the predominance of CO over CH$_4$ as the main form of carbon \citep{Hu15}, which could fit with the overall bulk composition of the planet determined here.

Another seemingly probable scenario, considering the planets' proximity to the star and the implied intense insolation, coupled with an assumed high water content of both planets, could be that the observed radii are highly inflated as the atmospheres may be well-represented by supercritical hydrospheres \citep{Mousis2020}. 
Unfortunately, the transmission spectroscopy metrics \citep[TSM,][]{Kempton18} for \planetb\ and \planetc\  are 44 and 42, respectively. This places both planets below the recommended TSM cutoff for planets with radii above 1.5~\rearth\ (TSM~>~90). Still, ground-based high-resolution spectroscopy could probe for the presence of ongoing escape processes by observing the \halpha\ lines \citep{Yan18} in the near-IR, as the \lyalpha\ line will be too absorbed by the interstellar medium at the system's distance ($\sim$~74~pc).




\section{Conclusions}

\label{sec:conclusions}

In this paper we present the detection and characterization of the \target\ system observed by TESS in Sectors 14 and 21. This \stype\ star hosts two mini-Neptunes in short-period orbits confirmed by HARPS-N radial velocities, as well as a tentative outer planet, which is seen transiting in the TESS photometry in Sector 21.

We use GP regression to disentangle the stellar from the planet signals contained in our radial velocities. GPs offer a lot of flexibility, which may lead to the removal of genuine signals of planetary origin - a risk we mitigate by using the information provided by activity indicators, i.e. the relatively novel multi-dimensional GP approach.

We note, however, that in order to improve the mass characterisation of the planets we need a strategic RV follow-up. More specifically, taking several observations within a single stellar rotation period, instead of sporadic observations, is a better strategy to disentangle stellar activity using GPs, since the latter rely on the correlation between points. 

We perform simulations to evaluate the possibility of hydrodynamic atmospheric mass loss, which demonstrated the difficulty in constraining the structure and composition of planets in 2 -- 3 \rearth\ radius range. Our discussion thus emphasizes the fact that solely from the mass and radius we cannot distinguish between a planet being H$_2$O-dominated or a rocky planet with a significant envelope. Another constraint to our insight into similar systems is the large uncertainty on the systems' ages. This could be remedied from a large sample of planet systems with well-determined ages, such as is attempted to be achieved by the core sample of the PLATO mission \citep{2014ExA....38..249R}, with projected uncertainties in its age determinations to be within 10\%.
In this paper we further demonstrate the need to study close-in planets around low-mass stars to help constrain composition models and mass-loss mechanisms. We add that the precision to which planetary masses are measured today is often insufficient to accomplish this to a satisfactory level, complicating our overall understanding of exoplanet demographics.

\section*{Acknowledgements}
This work is done under the framework of the KESPRINT collaboration 
(http://kesprint.science). KESPRINT is an international consortium devoted to the characterization and research of exoplanets discovered with space-based missions.

IYG, CMP, MF and JK gratefully acknowledge the support of the  Swedish National Space Agency (DNR 174/18, 65/19, 2020-00104).

KWFM and ME  acknowledge   the   support   of   the  DFG priority   program SPP  1992  "Exploring  the  Diversity of  Extrasolar  Planets" (RA714/14-1, HA3279/12-1).

HD acknowledges support from the Spanish Research Agency of the Ministry of Science and Innovation (AEI-MICINN) under grant PID2019-107061GB-C66, DOI: 10.13039/501100011033.

This work was supported by the Th\"uringer Ministerium f\"ur
Wirtschaft, Wissenschaft und Digitale Gesellschaft.

This research has made use of the Exoplanet Follow-up Observation Program website, which is operated by the California Institute of Technology, under contract with the National Aeronautics and Space Administration under the Exoplanet Exploration Program.

This work makes use of observations from the LCOGT network.

Some of the Observations in the paper made use of the High-Resolution Imaging instrument ‘Alopeke. ‘Alopeke was funded by the NASA Exoplanet Exploration Program and built at the NASA Ames Research Center by Steve B. Howell, Nic Scott, Elliott P. Horch, and Emmett Quigley. ‘Alopeke is mounted on the Gemini North telescope of the international Gemini Observatory, a program of NOIRLab, which is managed by the Association of Universities for Research in Astronomy (AURA) under a cooperative agreement with the National Science Foundation. on behalf of the Gemini partnership: the National Science Foundation (United States), National Research Council (Canada), Agencia Nacional de Investigación y Desarrollo (Chile), Ministerio de Ciencia, Tecnología e Innovación (Argentina), Ministério da Ciência, Tecnologia, Inovações e Comunicações (Brazil), and Korea Astronomy and Space Science Institute (Republic of Korea).

This paper includes data collected by the \tess{} mission. Funding for the \tess{} mission is provided by the NASA Explorer Program. We acknowledge the use of public TOI Release data from pipelines at the \tess{} Science Office and at the \tess{} Science Processing Operations Center. Resources supporting this work were provided by the NASA High-End Computing (HEC) Program through the NASA Advanced Supercomputing (NAS) Division at Ames Research Center for the production of the SPOC data products. This research has made use of the Exoplanet Follow-up Observation Program website, which is operated by the California Institute of Technology, under contract with the National Aeronautics and Space Administration under the Exoplanet Exploration Program.

This work has made use of data from the European Space Agency (ESA) mission {\it Gaia} (\url{https://www.cosmos.esa.int/gaia}), processed by the {\it Gaia} Data Processing and Analysis Consortium (DPAC, \url{https://www.cosmos.esa.int/web/gaia/dpac/consortium}). Funding for the DPAC has been provided by national institutions, in particular the institutions participating in the {\it Gaia} Multilateral Agreement.

This research has made use of the VizieR catalogue access tool, CDS, Strasbourg, France (DOI: 10.26093/cds/vizier). The original description of the VizieR service was published in A\&AS 143, 23.

Based on observations made with the Italian Telescopio Nazionale Galileo (TNG) operated on the island of La Palma by the Fundaci\'on Galileo Galilei of the INAF (Istituto Nazionale di Astrofisica) at the Spanish Observatorio del Roque de los Muchachos of the Instituto de Astrofisica de Canarias under programmes CAT19A\_162, ITP19\_1 and A41TAC\_49.


\noindent {\it Facility:} \tess, \gaia, TNG/HARPS-N, LCOGT.

\noindent {\it Software:} \texttt{EXOTRANS}, \texttt{lightkurve}, \texttt{citlalicue}, \texttt{george}, \texttt{pytransit}, \texttt{IRAF}, \texttt{PARAM~1.5}, \texttt{SME}, \texttt{SpecMatch-emp}, \texttt{AstroImageJ}, \texttt{pyaneti}, \texttt{juliet}, \texttt{pyTTV}.

\noindent \emph{Data availability:} The data underlying this article are available in the article and in its online supplementary material, as well as ExoFOP-TESS\footnote{https://exofop.ipac.caltech.edu/tess}.




\bibliographystyle{mnras}
\bibliography{bib}



\section*{List of affiliations}

$^{1}$Department of Space, Earth and Environment,\\
Chalmers University of Technology, Onsala Space Observatory, 439 92 Onsala, Sweden\\
$^{2}$Sub-department of Astrophysics, Department of Physics, University of Oxford, Oxford, OX1 3RH, UK\\
$^{3}$Instituto de Astrof\'{i}sica de Canarias, 38205 La Laguna, Tenerife, Spain\\
$^{4}$Departamento de Astrof\'{i}sica, Universidad de La Laguna, 38206 La Laguna, Tenerife, Spain\\
$^{5}$Leiden Observatory, Leiden University, 2333CA Leiden, The Netherlands\\
$^{6}$INAF - Osservatorio Astronomico di Palermo, Piazza del Parlamento 1, 90134 Palermo, Italy\\
$^{7}$Astronomy Department and Van Vleck Observatory, Wesleyan University, Middletown, CT 06459, USA\\
$^{8}$Dipartimento di Fisica, Universit\`{a} di Torino, via P. Giuria 1, 10125 Torino, Italy\\
$^{9}$Department of Earth and Planetary Sciences, University of California, Riverside, CA 92521, USA\\
$^{10}$Department of Space, Earth and Environment, Astronomy and Plasma Physics, Chalmers University of Technology, 412 96 Gothenburg, Sweden\\
$^{11}$Vanderbilt University, Physics and Astronomy Department, Nashville, TN 37235, USA\\
$^{12}$Department of Astronomy, University of Tokyo, 7-3-1 Hongo, Bunkyo-ku, Tokyo 113-0033, Japan\\
$^{13}$Department of Physics and Kavli Institute for Astrophysics and Space Research, Massachusetts Institute of Technology, Cambridge, MA 02139, USA\\
$^{14}$Observatoire de l’Universit\'e de Gen\`eve, Chemin des Maillettes 51, 1290 Versoix, Switzerland\\
$^{15}$Center for Astrophysics \textbar \ Harvard \& Smithsonian, 60 Garden Street, Cambridge, MA 02138, USA\\
$^{16}$NASA Ames Research Center, Moffett Field, CA 94035, USA\\
$^{17}$Stellar Astrophysics Centre, Department of Physics and Astronomy, Aarhus University, Ny Munkegade 120, DK-8000 Aarhus C, Denmark\\
$^{18}$Nordic Optical Telescope, Apartado 474, E-38700 Santa Cruz de La Palma, Santa Cruz de Tenerife, Spain\\
$^{19}$Space Telescope Science Institute, 3700 San Martin Dr, Baltimore, MD 21218, USA\\
$^{20}$NASA Exoplanet Science Institute – Caltech/IPAC Pasadena, CA 91125 USA\\
$^{21}$Center for Planetary Systems Habitability and McDonald Observatory, The University of Texas, Austin TX USA
$^{22}$NASA Goddard Space Flight Center, Exoplanets and Stellar Astrophysics Laboratory (Code 667), Greenbelt, MD 20771, USA\\
$^{23}$Department of Physics and Astronomy, University of Kansas, Lawrence, KS, USA\\
$^{24}$Institute of Planetary Research, German Aerospace Center (DLR), Rutherfordstraße 2, 12489 Berlin, Germany\\
$^{25}$Thüringer  Landessternwarte  Tautenburg,
Sternwarte  5,  D-07778Tautenberg, Germany\\
$^{26}$NASA Exoplanet Science Institute, Caltech/IPAC, Mail Code 100-22, 1200 E. California Blvd., Pasadena, CA 91125, USA\\
$^{27}$Department of Astronomy and Tsinghua Centre for Astrophysics, Tsinghua University, Beijing 100084, China\\
$^{28}$Department of Astronomy and Astrophysics, University of California, Santa Cruz, 1156 High St. Santa Cruz , CA 95064, USA\\
$^{29}$Rheinisches Institut für Umweltforschung an der Universität zu Köln, Aachener Strasse 209, 50931 Köln
$^{30}$Observatori Astronòmic Albanyà, Camí de Bassegoda S/N, Albanyà 17733, Girona, Spain\\
$^{31}$Astrobiology Center, NINS, 2-21-1 Osawa, Mitaka, Tokyo 181-8588, Japan\\
$^{32}$National Astronomical Observatory of Japan, NINS, 2-21-1 Osawa, Mitaka, Tokyo 181-8588, Japan\\
$^{33}$Department of Physics \& Astronomy, Swarthmore College, Swarthmore PA 19081, USA\\
$^{34}$Astronomical Institute ASCR, Fri{\v c}ova 298,251 65, Ond{\v r}ejov, Czech Republic\\
$^{35}$Center for Astronomy and Astrophysics, Technical University Berlin, Hardenbergstr. 36, 10623 Berlin, Germany\\
$^{36}$U.S. Naval Observatory, Washington, D.C. 20392, USA\\
$^{37}$Proto-Logic LLC, 1718 Euclid Street NW, Washington, DC 20009, USA\\
$^{38}$Mullard Space Science Laboratory, University College London, Holmbury St Mary, Dorking, Surrey, RH5 6NT, UK\\
$^{39}$NASA Goddard Space Flight Center, 8800 Greenbelt Rd., Greenbelt, MD 20771, USA\\
$^{40}$Department of Earth, Atmospheric and Planetary Sciences, Massachusetts Institute of Technology, Cambridge, MA 02139, USA\\
$^{41}$Department of Aeronautics and Astronautics, MIT, 77 Massachusetts Avenue, Cambridge, MA 02139, USA\\
$^{42}$SETI Institute, Mountain View, CA 94043, USA\\
$^{43}$Department of Astrophysical Sciences, Princeton University, 4 Ivy Lane, Princeton, NJ 08544, USA\\

\appendix

\section{HARPS-N data}
\null\newpage
\clearpage
\input{all_rv.tex}

\section{Monotransit periods}
\null\newpage
\clearpage

\begin{table}
\caption{Possible periods and period ranges for the case of a unique single transit assuming a circular orbit. The excluded values/gaps correspond to transit times when a transit event would be seen in the light curves. The calculations were performed in steps of 0.1 days and include the data gap in Sector 14 as a possible location of a missed transit. The table does not include the 16.6-day period corresponding to the scenario of overlapping transits of this tentative planet and planet c, described in the text. If this period is correct, this would imply the presence of two transits in Sector 21.
\label{mono_P}}
\begin{tabular}{l}
\hline
\noalign{\smallskip}
Period (days) \\
\noalign{\smallskip}
     \hline
\noalign{\smallskip} 
20.3        \\
22.8        \\
26.0 - 26.1 \\
28.0 - 28.1 \\
30.4        \\
32.7 - 33.8 \\
36.4 - 36.5 \\
39.2 - 42.2 \\
45.5 - 45.7 \\
49.0 - 56.3 \\
\noalign{\smallskip}
\hline
\end{tabular}
\end{table}


\bsp	
\label{lastpage}
\end{document}

%% file: all_rv.tex
\begin{sidewaystable}
\begin{ssmall}
\begin{center}
  \caption{Radial velocities and spectral activity indicators measured from TNG/HARPS-N spectra.
    \label{all_rv.tex}}
  \begin{tabular}{rrrrrrrrrrrrrrrrrrrrr}
    \hline
    \hline
    \multicolumn{1}{r}{BJD$_\mathrm{TBD}$ (days)} &
    \multicolumn{1}{r}{RV} &
    \multicolumn{1}{r}{$\sigma_\mathrm{RV}$} &
    \multicolumn{1}{r}{CRX} &
    \multicolumn{1}{r}{$\sigma_\mathrm{CRX}$} &
    \multicolumn{1}{r}{dlW} &
    \multicolumn{1}{r}{$\sigma_\mathrm{dlW}$} &
    \multicolumn{1}{r}{$\mathrm{H_{\alpha}}$} &
    \multicolumn{1}{r}{$\mathrm{\sigma_{H_{\alpha}}}$} &
    \multicolumn{1}{r}{$\mathrm{NaD_{1}}$} &
    \multicolumn{1}{r}{$\mathrm{\sigma_{NaD_{1}}}$} &
    \multicolumn{1}{r}{$\mathrm{NaD_{2}}$} &
    \multicolumn{1}{r}{$\mathrm{\sigma_{NaD_{2}}}$} &
    \multicolumn{1}{r}{BIS} &
    \multicolumn{1}{r}{$\sigma_\mathrm{BIS}$} &
    \multicolumn{1}{r}{CCF\_FHWM} &
    \multicolumn{1}{r}{CCF\_CTR} &
    \multicolumn{1}{r}{S-index} &
    \multicolumn{1}{r}{$\sigma_\mathrm{S-index}$} &
    \multicolumn{1}{r}{SNR} &
    \multicolumn{1}{r}{$\mathrm{T_{exp}}$}\\ 
    \multicolumn{1}{r}{-2457000} &
    \multicolumn{1}{r}{($\mathrm{m\,s^{-1}}$)} &
    \multicolumn{1}{r}{($\mathrm{m\,s^{-1}}$)} &
    \multicolumn{1}{r}{($\mathrm{m\,s^{-1}\,Np^{-1}}$)} &
    \multicolumn{1}{r}{($\mathrm{m\,s^{-1}\,Np^{-1}}$)} &
    \multicolumn{1}{r}{($\mathrm{m^2\,s^{-2}}$)} &
    \multicolumn{1}{r}{($\mathrm{m^2\,s^{-2}}$)} &
    \multicolumn{1}{r}{---} &
    \multicolumn{1}{r}{---} &
    \multicolumn{1}{r}{---} &
    \multicolumn{1}{r}{---} &
    \multicolumn{1}{r}{---} &
    \multicolumn{1}{r}{---} &
    \multicolumn{1}{r}{(\mps)} &
    \multicolumn{1}{r}{(\mps)} &
    \multicolumn{1}{r}{(\kmps)} &
    \multicolumn{1}{r}{---} &
    \multicolumn{1}{r}{---} &
    \multicolumn{1}{r}{---} &
    \multicolumn{1}{r}{---} &
    \multicolumn{1}{r}{(s)}\\ 
    \hline
     1862.62923 &           6.677 &           1.716 &         -17.725 &          14.096 &          -2.805 &           3.196 &          0.6841 &          0.0020 &          0.2004 &          0.0024 &          0.2252 &          0.0026 &          40.628 &           3.763 &           6.364 &          42.479 &           1.133 &           0.025 &            38.7 &          3600.0\\
     1862.66579 &           4.803 &           1.340 &          -9.964 &          11.112 &          -7.302 &           2.351 &          0.6850 &          0.0017 &          0.1974 &          0.0020 &          0.2216 &          0.0021 &          50.620 &           3.022 &           6.361 &          42.637 &           1.122 &           0.019 &            45.9 &          3600.0\\
     1863.67080 &          10.163 &           1.235 &          10.432 &          10.103 &         -17.998 &           1.895 &          0.6832 &          0.0013 &          0.1945 &          0.0014 &          0.2200 &          0.0015 &          47.085 &           2.120 &           6.360 &          42.878 &           1.134 &           0.012 &            61.2 &          3600.0\\
     1863.78184 &          11.454 &           1.180 &          10.563 &           9.470 &         -15.791 &           1.870 &          0.6844 &          0.0012 &          0.1951 &          0.0014 &          0.2196 &          0.0015 &          52.086 &           2.173 &           6.363 &          42.821 &           1.135 &           0.012 &            59.9 &          3600.0\\
     1864.64888 &          15.901 &           1.490 &          14.765 &          12.360 &          -3.886 &           2.166 &          0.6855 &          0.0018 &          0.1975 &          0.0020 &          0.2236 &          0.0022 &          56.202 &           3.091 &           6.371 &          42.728 &           1.154 &           0.021 &            45.1 &          3600.0\\
     1864.76150 &          13.179 &           1.573 &          20.240 &          13.193 &          -6.655 &           2.239 &          0.6827 &          0.0019 &          0.1968 &          0.0021 &          0.2284 &          0.0023 &          46.279 &           3.192 &           6.363 &          42.705 &           1.168 &           0.022 &            44.2 &          3600.0\\
     1865.63156 &           5.489 &           1.193 &          14.321 &          10.311 &          -8.966 &           2.404 &          0.6816 &          0.0017 &          0.1991 &          0.0016 &          0.2245 &          0.0018 &          49.886 &           2.354 &           6.380 &          42.741 &           1.141 &           0.014 &            56.3 &          3600.0\\
     1865.76718 &           1.903 &           1.434 &          19.806 &          11.892 &          -8.575 &           2.433 &          0.6913 &          0.0016 &          0.2000 &          0.0016 &          0.2219 &          0.0018 &          47.921 &           2.470 &           6.381 &          42.767 &           1.169 &           0.016 &            54.5 &          3600.0\\
     1869.71163 &           0.562 &           1.555 &          -9.033 &          12.511 &          -1.288 &           2.907 &          0.6875 &          0.0017 &          0.2007 &          0.0021 &          0.2282 &          0.0023 &          63.711 &           3.314 &           6.383 &          42.718 &           1.193 &           0.022 &            41.9 &          3600.0\\
     1896.56585 &          15.450 &           1.436 &          -3.060 &          11.637 &          -1.510 &           2.297 &          0.6878 &          0.0016 &          0.2237 &          0.0020 &          0.2459 &          0.0021 &          45.665 &           2.994 &           6.384 &          42.720 &           1.214 &           0.019 &            45.9 &          3600.0\\
     1896.67552 &          13.918 &           1.388 &         -11.290 &          11.143 &          -0.808 &           2.814 &          0.6945 &          0.0016 &          0.2284 &          0.0020 &          0.2467 &          0.0022 &          43.192 &           3.018 &           6.386 &          42.659 &           1.232 &           0.019 &            45.8 &          3600.0\\
     1898.56841 &           7.774 &           1.073 &          -3.532 &           8.928 &           0.857 &           2.620 &          0.7382 &          0.0014 &          0.2165 &          0.0015 &          0.2407 &          0.0016 &          56.588 &           2.184 &           6.387 &          42.631 &           1.446 &           0.013 &            60.1 &          3600.0\\
     1925.63670 &          17.927 &           1.749 &          -2.160 &          14.638 &          -1.086 &           3.147 &          0.6863 &          0.0023 &          0.2479 &          0.0030 &          0.2599 &          0.0032 &          43.613 &           4.335 &           6.393 &          42.655 &           1.123 &           0.032 &            34.8 &          3600.0\\
     1926.64886 &          19.622 &           2.118 &         -15.619 &          17.407 &           0.278 &           2.686 &          0.7005 &          0.0019 &          0.2343 &          0.0025 &          0.2600 &          0.0027 &          61.432 &           3.705 &           6.379 &          42.775 &           1.199 &           0.026 &            39.4 &          3300.0\\
     1929.63674 &          10.128 &           1.193 &          -1.733 &          10.068 &          -7.456 &           2.374 &          0.6858 &          0.0018 &          0.2118 &          0.0020 &          0.2432 &          0.0022 &          64.774 &           2.971 &           6.372 &          42.772 &           1.175 &           0.020 &            47.1 &          3006.7\\
     1930.42198 &          11.743 &           2.290 &         -24.218 &          19.108 &          -6.858 &           3.472 &          0.6937 &          0.0028 &          0.2339 &          0.0036 &          0.2459 &          0.0039 &          72.455 &           5.391 &           6.359 &          42.761 &           1.026 &           0.054 &            29.6 &          2200.0\\
     1969.53394 &          -0.344 &           2.607 &          30.922 &          21.458 &         -23.205 &           3.971 &          0.6870 &          0.0031 &          0.3179 &          0.0047 &          0.3119 &          0.0049 &          52.253 &           6.453 &           6.339 &          42.808 &           1.146 &           0.041 &            25.9 &          3600.0\\
     1971.52051 &          -2.375 &           1.298 &          -6.884 &          10.783 &         -44.328 &           2.698 &          0.6768 &          0.0016 &          0.2088 &          0.0019 &          0.2320 &          0.0021 &          63.058 &           2.849 &           6.316 &          43.145 &           1.011 &           0.019 &            48.1 &          3600.0\\
     2000.41367 &           2.216 &           1.268 &          -8.945 &          10.158 &         -36.375 &           2.377 &          0.6827 &          0.0014 &          0.2021 &          0.0017 &          0.2323 &          0.0019 &          70.680 &           2.731 &           6.333 &          43.037 &           1.098 &           0.018 &            48.8 &          3600.0\\
     2000.45426 &           4.378 &           1.489 &         -35.801 &          10.971 &         -25.136 &           2.493 &          0.6822 &          0.0015 &          0.2101 &          0.0021 &          0.2300 &          0.0022 &          75.663 &           3.223 &           6.333 &          43.001 &           1.067 &           0.022 &            42.4 &          3600.0\\
     2001.42014 &           0.440 &           1.116 &         -11.500 &           8.915 &         -47.568 &           1.854 &          0.6692 &          0.0011 &          0.1937 &          0.0012 &          0.2238 &          0.0014 &          63.516 &           1.940 &           6.312 &          43.220 &           1.038 &           0.010 &            65.8 &          3600.0\\
     2001.46039 &          -0.236 &           0.949 &         -23.011 &           7.095 &         -48.208 &           1.735 &          0.6705 &          0.0010 &          0.1916 &          0.0011 &          0.2242 &          0.0012 &          73.278 &           1.702 &           6.312 &          43.235 &           1.024 &           0.009 &            73.8 &          3600.0\\
     2002.42024 &          -2.632 &           0.905 &           5.811 &           7.749 &         -53.132 &           2.055 &          0.6726 &          0.0014 &          0.1975 &          0.0014 &          0.2341 &          0.0016 &          56.605 &           1.996 &           6.303 &          43.312 &           1.001 &           0.011 &            64.2 &          3600.0\\
     2002.46195 &          -0.506 &           1.311 &          13.103 &          10.881 &         -55.646 &           2.419 &          0.6712 &          0.0013 &          0.1967 &          0.0014 &          0.2353 &          0.0015 &          63.482 &           1.999 &           6.302 &          43.323 &           1.024 &           0.011 &            64.4 &          3600.0\\
     2011.40922 &          27.029 &           2.205 &          -0.278 &          18.301 &         -29.220 &           3.238 &          0.6777 &          0.0024 &          0.1989 &          0.0033 &          0.2373 &          0.0036 &          48.672 &           5.219 &           6.321 &          43.031 &           1.123 &           0.041 &            30.2 &          3300.0\\
     2011.44218 &          23.398 &           2.840 &          16.880 &          23.534 &         -32.558 &           4.204 &          0.6794 &          0.0034 &          0.2268 &          0.0054 &          0.2724 &          0.0060 &          54.360 &           8.605 &           6.334 &          42.874 &           1.069 &           0.075 &            20.8 &          2999.1\\
     2012.40959 &          25.530 &           1.828 &          20.100 &          14.729 &         -24.200 &           2.180 &          0.6723 &          0.0016 &          0.2009 &          0.0020 &          0.2305 &          0.0022 &          33.998 &           3.095 &           6.349 &          42.922 &           1.105 &           0.021 &            45.2 &          3300.0\\
     2012.44888 &          25.937 &           1.907 &          13.461 &          15.545 &         -27.838 &           2.505 &          0.6822 &          0.0018 &          0.2038 &          0.0022 &          0.2337 &          0.0024 &          45.139 &           3.518 &           6.342 &          42.933 &           1.157 &           0.026 &            40.7 &          3300.0\\
     2013.42280 &          17.344 &           1.733 &          30.803 &          13.985 &         -20.223 &           2.502 &          0.6878 &          0.0020 &          0.2484 &          0.0025 &          0.2671 &          0.0026 &          35.178 &           3.499 &           6.354 &          42.858 &           1.152 &           0.024 &            41.5 &          3300.0\\
     2013.44892 &          17.566 &           2.332 &           0.142 &          20.049 &         -13.318 &           4.085 &          0.6896 &          0.0035 &          0.2526 &          0.0049 &          0.2816 &          0.0054 &          54.908 &           7.476 &           6.371 &          42.777 &           1.121 &           0.063 &            23.7 &          1353.0\\
     2013.47580 &          26.451 &           3.139 &          39.435 &          26.754 &          -9.109 &           5.318 &          0.6887 &          0.0042 &          0.3239 &          0.0064 &          0.3563 &          0.0069 &          14.612 &           8.586 &           6.351 &          42.772 &           1.103 &           0.075 &            21.0 &          2273.7\\
     2014.42732 &          30.021 &           1.846 &          -7.137 &          15.000 &         -13.315 &           2.384 &          0.6837 &          0.0017 &          0.2265 &          0.0024 &          0.2527 &          0.0025 &          46.902 &           3.644 &           6.370 &          42.766 &           1.168 &           0.027 &            39.5 &          3300.0\\
     2014.46531 &          28.500 &           1.704 &          16.641 &          13.691 &         -14.391 &           3.104 &          0.6796 &          0.0018 &          0.2181 &          0.0024 &          0.2506 &          0.0026 &          39.564 &           3.733 &           6.360 &          42.848 &           1.165 &           0.029 &            39.0 &          3300.0\\
    \hline
  \end{tabular}
\end{center}
\end{ssmall}
\end{sidewaystable}